\documentclass[paper=A4,11pt,DIV=12,headings=big]{scrartcl}
\pdfoutput=1

\usepackage[a4paper, top=1.2in, bottom=1.2in, left=1.1in, right=1.1in]{geometry}
\usepackage{amsfonts}
\usepackage{amsmath,amssymb,slashed}
\usepackage[normal]{caption}

\usepackage{cancel}
\usepackage{xcolor}
\usepackage{cite,bbm}
\usepackage{enumerate}
\usepackage{graphicx}
\graphicspath{{./Figs/}}
\usepackage{wrapfig}

\renewenvironment{abstract}{%
\begin{minipage}{0.95\textwidth}
}
{\par\noindent\end{minipage}}

\usepackage[multiple]{footmisc}
\let\oldfootnote\footnote\renewcommand\footnote[1]{\oldfootnote{\hspace{2mm}#1}}

\definecolor{darkblue}{rgb}{0,0,0.9}
\usepackage{hyperref}
\hypersetup{
linktocpage,
colorlinks,
citecolor=darkblue,
filecolor=darkblue,
linkcolor=darkblue,
urlcolor=darkblue,
pdftitle={MT2-reconstructed invisible momenta as spin analizers, and an application to top polarization},%
pdfauthor={Diego Guadagnoli, Chan Beom Park}
}
\allowdisplaybreaks
\addtokomafont{disposition}{\rmfamily\boldmath}
\newcommand{\mailref}[1]{\href{mailto:#1}{#1}}

\def\MPT{\slashed{\boldsymbol{p}}_{\rm T}}
\def\T{{\rm T}}
\def\mt{M_{\rm T}}
\def\MT2{M_{\rm T2}}
\def\k#1{{\boldsymbol{k}}_{#1}}
\def\kT#1{{\boldsymbol{k}}_{#1 {\rm T}}}
\def\okT#1{\bar{\boldsymbol{k}}_{#1 {\rm T}}}
\def\kL#1{{\bar k}_{#1 {\rm L}}^{\pm}}
\def\p#1{{\boldsymbol{p}}_{#1}}

\def\maos{{\rm maos}}
\def\true{{\rm true}}
\def\max{{\rm max}}
\newcommand{\ttbar}{t \bar t}
\newcommand{\tutu}{t_\uparrow \bar t_\uparrow}
\newcommand{\tdtd}{t_\downarrow \bar t_\downarrow}
\newcommand{\tutd}{t_\uparrow \bar t_\downarrow}
\newcommand{\tdtu}{t_\downarrow \bar t_\uparrow}

\def\sla#1{\setbox0=\hbox{$#1$}\dimen0=\wd0
      \setbox1=\hbox{/} \dimen1=\wd1 \ifdim\dimen0>\dimen1
      \rlap{\hbox to \dimen0{\hfil/\hfil}} #1                        \else
      \rlap{\hbox to \dimen1{\hfil$#1$\hfil}}
      /   \fi}

\newcommand{\be}{\begin{equation}}
\newcommand{\ee}{\end{equation}}
\newcommand{\bea}{\begin{eqnarray}}
\newcommand{\eea}{\end{eqnarray}}

\begin{document}

%%%% TITLE PAGE
\begin{flushright}
\small
LAPTH-042/13\\
CERN-PH-TH/2013-190
\end{flushright}
\vskip0.5cm

%\begin{flushright}
%Last modified on \today
%\end{flushright}

\begin{center}
%%%% TITLE
{\sffamily \bfseries \LARGE \boldmath
$\MT2$-reconstructed invisible momenta as spin analizers,\\[0.2cm]
and an application to top polarization}\\[0.8 cm]
%%%% AUTHORS
{\normalsize \sffamily \bfseries Diego~Guadagnoli$^a$ and Chan~Beom~Park$^{b}$} \\[0.5 cm]
\small
$^a$ {\em LAPTh, Universit\'e de Savoie et CNRS, BP110, F-74941 Annecy-le-Vieux Cedex, France}\\[0.1cm]
$^b${\em CERN, Theory Division, 1211 Geneva 23, Switzerland}\\[0.4cm]
{\em E-mail:} \mailref{diego.guadagnoli@lapth.cnrs.fr}, \mailref{chanbeom.park@cern.ch}%
\end{center}

\medskip

\begin{abstract}\noindent
Full event reconstruction is known to be challenging in cases with more than one undetected
final-state particle, such as pair production of two states each decaying semi-invisibly.
On the other hand, full event reconstruction would allow to access angular distributions 
sensitive to the spin fractions of the decaying particles, thereby dissecting their 
production mechanism.
We explore this possibility in the case of Standard-Model $\ttbar$ production followed by
a leptonic decay of both $W$ bosons, implying two undetected final-state neutrinos. We 
estimate the $t$ and $\bar t$ momentum vectors event by event using information extracted 
from the kinematic variable $\MT2$. The faithfulness of the estimated momenta to the true momenta 
is then tested in observables sensitive to top polarization and $\ttbar$ spin correlations. 
Our method thereby provides a novel approach towards the evaluation of these observables, and 
towards testing $\ttbar$ production beyond the level of the total cross section.
While our discussion is confined to $\ttbar$ production as a benchmark, the method is applicable
to any process whose decay topology allows to construct $\MT2$.
\end{abstract}

\vspace{1.0cm}

%%%% TOC
\setcounter{tocdepth}{2}
\noindent \rule{\textwidth}{0.3pt}\vspace{-0.4cm}\tableofcontents
\noindent \rule{\textwidth}{0.3pt}
%%%% END TOC

\renewcommand{\thefootnote}{\arabic{footnote}}
\setcounter{footnote}{0}
%%%% END TITLE PAGE

\section{Introduction}

\noindent Top-quark decays are known to be privileged places for testing the
Standard Model (SM) and for providing hints on the theory that completes
it at high energies. There are theoretical, phenomenological and experimental
reasons for this fact. At the theory level, within the SM the top quark
is the only `heavy' fermion, with namely mass of the order of the electroweak
symmetry breaking (EWSB) scale,
or equivalently with Higgs coupling of $\mathcal{O}(1)$. This circumstance 
motivates beyond-SM scenarios where the top quark plays an active role in
the EWSB dynamics, at variance with the other quarks.
At the phenomenological level, the large top mass causes the top quark
to decay before hadronization, so that the details of its production
mechanism (e.g. the relative weights of the different spin amplitudes)
are testable from the kinematic behavior of its decay products.
Finally, at the experimental level, any collider experiment devoted to
directly exploring the EWSB scale is in principle also a top-quark factory.
Most notably, this is true for the LHC that, in 2011 alone, has produced
as many as $8 \cdot 10^5$ $t \bar t$ pairs per experiment \cite{Schilling:2012dx}.

Within the SM, the top quark decays almost exclusively as $t \to W b$.
Therefore, the final states are the same as those of the $W$ boson, aside
from an additional $b$ jet. Final states in $t \bar t$ can accordingly be classified
as fully hadronic, semi-leptonic, and di-leptonic, depending on the decay
modes of the two $W$ bosons. Among them, the di-leptonic final state, consisting
of two $b$ jets, two charged leptons, and two neutrinos, is of particular
interest: it provides a clean signal because of the charged leptons, and
its topology (pair production of visibles plus missing energy)
resembles the typical signatures of beyond-SM models with a dark
matter candidate made stable by some discrete parity.

Full event reconstruction in di-leptonic $\ttbar$ decays
poses a challenge because of the two undetected neutrinos. This is especially
true at hadron colliders. In fact, theoretically the parent four-momenta
can be calculated analytically, as was shown in Ref.~\cite{Sonnenschein:2006ud},
because the six available kinematic constraints (invariant masses for
$t$, $\bar t$, $W^+$, $W^-$ and transverse missing momentum) are enough
to solve for the six unknowns (the $t \bar t$ momenta, or equivalently the
neutrino momenta, assuming a perfect measurement of the visibles' momenta).
However, in real life finite detector resolutions and the imperfect, sometimes
poor, particle identification result in a proliferation of the actual number of
analytic solutions \cite{Sonnenschein:2006ud}, making this method impractical.
This leads the experimental collaborations to either opt for maximum-likelihood-inspired 
methods, or else to resort to observables defined in the lab frame.

In this paper we explore the possibility of reconstructing the full $t$
and $\bar t$ boosts in di-leptonic $t \bar t$ decays, using information
extracted from the kinematic variable $\MT2$ \cite{Lester:1999tx,Barr:2003rg}.
As well known, lack of knowledge of the $t$ and $\bar t$ momenta impairs 
evaluation of several top-polarization and $t \bar t$ spin-correlations observables. 
We calculate these observables with the $t$ and $\bar t$ momenta determined with our 
approach. Our results make the underlying method a potential novel avenue towards the 
measurement of these observables in di-leptonic $\ttbar$ decays.

The $\MT2$ variable is the pair-production generalization of the $\mt$
variable \cite{Smith:1983aa}, extensively used e.g. for $W$ mass
measurements in $W \to \ell \nu$. This decay is the simplest decay
to a visible plus an invisible particle. In the notation of this decay,
the $\mt$ variable reads
\be
\mt^2 = m_\ell^2 + m_\nu^2 + 2 (E^\ell_{\T} E^\nu_{\T} -
\p{\T}^\ell \p{\T}^\nu)~.
\label{eq:MTdef}
\ee
The same expression, with the $E^\ell_{\T} E^\nu_{\T}$ factor multiplied
by $\cosh(\eta_\ell - \eta_\nu)$ and $\eta$ the particle rapidity, would
equal $m_W^2$. Therefore $\mt \leq m_W$. Since kinematic configurations exist
for the equality to be fulfilled, the $\mt$ endpoint allows indeed to measure
the $W$ mass.
The generalization of this argument to two decay chains yields $\MT2$
\cite{Lester:1999tx,Barr:2003rg} as mentioned. The latter can be defined as follows
\be
  \MT2 \equiv
  \min_{\k{\T}^{(1)} + \k{\T}^{(2)} =
    \MPT} \left [
    \max \left \{
      M_\T \left( \p{\T}^{(1)}, \, \k{\T}^{(1)} \right), \,
      M_\T \left( \p{\T}^{(2)}, \, \k{\T}^{(2)} \right)
      \right \} \right ],
  \label{eq:MT2def}
\ee
with $p^{(i)}$ and $k^{(i)}$ $(i = 1,\,2)$ denoting the sum of the visible-particle
momenta and respectively the momentum of the undetected particle in either of the two decay 
chains, labelled by $i$.
Similarly as $\mt$, $\MT2$ provides, event by event, a lower bound on the mother
particle mass: in the case of di-leptonic $\ttbar$ decays $\MT2 \leq m_t$. As a matter of 
fact, $\MT2$ has been extensively used for mass measurements such as the top quark's,\footnote{%
For the original proposal in the context of the LHC, see \cite{Cho:2008cu}.}
both at Tevatron~\cite{Aaltonen:2009rm} and at the LHC~\cite{ATLAS:2012poa}.

However, this variable has a much wider spectrum of potential applicability.
In particular, by its very definition~\cite{Lester:1999tx,Barr:2003rg}, it is
designed to make the most out of topologies involving two decay
chains, each consisting of an undetected\footnote{\label{foot:undetected}%
We note in this respect that, in evaluating $\MT2$, the `undetected' particle does not really 
need be so. One may assign a neatly reconstructed charged lepton or a jet to the invisible part 
of the decay by just including its transverse momentum in the missing-momentum budget.}
particle and one or more visible particles, topologies often encountered in beyond-SM
extensions, as mentioned.

The $\MT2$ potentialities can be understood from the two main complications
naturally encountered when going from eq. (\ref{eq:MTdef}) to (\ref{eq:MT2def}). 
First, the invisible-daughter mass is not necessarily known. In such cases, $\MT2$ 
is a function of this mass event by event. 
In fact, it has been pointed out that, even in the absence of knowledge of the
mother- and of the invisible-particle masses (to be indicated with $m_Y$ and $m_X$,
respectively), the inequality $\MT2(m_X) \leq m_Y$ is a necessary and sufficient
condition for the decay kinematics to be physical \cite{ChengHan}.\footnote{\label{foot:Cheng}%
Ref. \cite{ChengHan} presents a neat proof of this statement and a 
discussion of its implications. For further insights, see also \cite{Barr:2009jv}.
Before this literature, the idea of $\MT2$ as the boundary of the physical region
in the $(m_X, m_Y)$ plane had been used more or less implicitly in \cite{RossSerna,MAOS}.
} 
In other words, event momenta fulfilling this relation will correctly satisfy the available 
kinematic constraints.

The second complication/potentiality is the fact that the invisible particles'
transverse momenta, $\k{\T}^{(i)}$ in eq. (\ref{eq:MT2def}), are not measured
individually -- only their sum is. Therefore, when constructing $\mt$ for each of
the two decay chains, there is a two-dimensional parameter space, represented e.g.
by $\k{\T}^{(1)}$, out of which one has to pick up a value. While in principle the
choice of $\k{\T}^{(1)}$ is arbitrary, several kinematic considerations, that we
will not repeat here (see e.g. \cite{Lester:1999tx,MAOS}), suggest to take the 
$\k{\T}^{(1)}$ value that
yields the minimum for the largest between the two $\mt$. Indicating this choice as 
$\okT{}$, we see that, by construction, 
$\max\{M_{\T}^{(1)}, M_{\T}^{(2)}\}|_{\k{\T}^{(1)} = \okT{}}$ equals exactly $\MT2$
(cf. eq. (\ref{eq:MT2def})).
In other words, the $\MT2$ evaluation, event by event, comes with a well-defined assignment
for the {\em individual} invisible particles' momenta: $\k{\T}^{(1)} = \okT{}$,
$\k{\T}^{(2)} = \MPT - \okT{}$. This assignment has been shown \cite{MAOS}
to be normally distributed around the true invisible momenta, and to provide, for several
practical purposes, an effective `best guess' of the true momenta. Following Ref.
\cite{MAOS}, we will refer to the thus assigned invisible momenta as `$\MT2$-assisted
on-shell' (MAOS) momenta.

In this paper we explore the question whether the $t$ or $\bar t$ boost reconstructed 
from MAOS-determined invisible momenta is faithful enough to the real $t$ or $\bar t$ boost,
that it can be used to evaluate observables sensitive to the top spin. We find that 
MAOS-calculated distributions measuring top polarization and $\ttbar$ spin correlations 
have shapes and asymmetries always close to the ones obtained using the true top boosts, 
and that deviations can be systematically improved by just $\MT2$ cuts. 

The technique, to be discussed in the next sections, can be adapted to the measurement 
of the spin distributions of any new particles produced in pairs. A vast literature exists on this 
topic, that is impossible to acknowledge in full. References to which our approach is directly
applicable, or has been applied, include \cite{Spin_th_refs}. (This list does not include work 
referred to later on within specific contexts.) More generally, provided one can construct 
$\MT2$, our approach may be applied to {\em any} observable requiring reconstruction of the 
parent-particle's boost. Several examples thereof exist e.g. among observables related to the 
forward-backward asymmetry in $\ttbar$ production, for which recent literature is even vaster.

In beyond-SM generalizations the method's performance will depend on the nature, 
production rate, decay modes and backgrounds of the new particles in question. We leave this topic 
outside the scope of the present work, that as mentioned will be focused on SM $\ttbar$ production 
as a benchmark case.

\section{MAOS reconstruction of the top rest frame}

\subsection[$\MT2$ and MAOS momenta]{\boldmath $\MT2$ and MAOS momenta}

In order to make the discussion as self-contained as possible, it is worthwhile
to shortly reproduce the line of reasoning \cite{MAOS,Park:2011uz} leading to the
definition of MAOS momenta.

Consider the following decay -- our discussion will apply to any process with
the same topology
\bea
\label{eq:decay}
Y_1 + Y_2 \rightarrow V(p_1) X(k_1) + V(p_2) X(k_2)~,
\eea
where $Y_{1,\,2}$ are pair-produced particles, assumed to have a common mass $m_Y$,
$V(p_i)$ are a set of one {\em or more} visible particles with total momentum $p_i$,
$X(k_i)$ are two undetected particles with momentum $k_i$ and mass $m_X$.
Daughters labelled with $i = 1, \, 2$ are assumed to be the decay products of
$Y_{1,\,2}$, respectively. The di-leptonic $t \bar t$ decay
\bea
\label{eq:ttbar_dilep}
t + \bar t \rightarrow b ~ W^+(\to \ell^+ \nu) + \bar b ~ W^-(\to \ell^- \bar \nu)
\eea
is a SM prototype of the decay process in eq. (\ref{eq:decay}), $V(p_i)$ being
the two $b \ell$ pairs and $X(k_i)$ being the undetected neutrinos.

In the process (\ref{eq:decay}) the momenta $p_i$ are assumed to be measurable,
along with the transverse component of the total missing momentum, $\MPT$.
Our final task is to reconstruct the full $Y_i$ boosts, for which we need to
reconstruct $k_i$ individually. We can write the following on-shell equations
\bea
\label{eq:on-shell}
(p_i + k_i)^2 =m_Y^2~, ~~ k_i^2 = m_X^2~, ~~ \kT{1} + \kT{2} = \MPT~,
\eea
corresponding to six constraints. In the general case of eq. (\ref{eq:decay}),
the unknowns include, besides $\k{1}$ and $\k{2}$, also the masses $m_Y$ and $m_X$,
so that there is a 2-parameter space of solutions,\footnote{%
In the case of di-leptonic $t \bar t$ $m_Y$ and $m_X$ are known, hence the only
unknowns are the two neutrino 3-momenta. Therefore, {\em in principle} the full
kinematics can be solved analytically. In practice, as mentioned in the Introduction
and elucidated in \cite{Sonnenschein:2006ud}, imperfect knowledge of the measurable
quantities leads to a proliferation of the analytic solutions.}
that can be parameterized by $\kT{1}$. Once $\kT{1}$ is fixed as two real numbers,
the longitudinal $k_i$ components, $k_{i \rm L}$, can be determined from (\ref{eq:on-shell})
as the solutions of two quadratic equations. In general, there will be therefore a
two-fold ambiguity on either of the $k_{i \rm L}$ solutions, that we indicate as $\kL{i}$.
Most interestingly, the condition that the discriminants of the two quadratic equations
be both real can be written as \cite{MAOS}
\bea
\label{eq:mYgeMTMT}
m_Y \ge {\rm max} \left\{ M_\T^{(1)}, \, M_\T^{(2)} \right \}~,
\eea
where $M_\T^{(i)}$ is the transverse mass constructed for decay chain $i$. This relation
shows a certain kinship with the $\MT2$ definition, and in fact one can go further.
The discussion
so far holds independently of the $\kT{1}$ choice, that we have not yet specified.
In fact, the r.h.s. of eq. (\ref{eq:mYgeMTMT}) should be seen as a function of
$\kT{1}$. This in turn suggests that, in order for the inequality to be fulfilled for
the largest possible number of events, the most `conservative' choice of $\kT{1}$ is the
one that yields the minimum of the r.h.s. of eq. (\ref{eq:mYgeMTMT}), under the constraint
$\kT{1} + \kT{2} = \MPT$. By comparing with eq. (\ref{eq:MT2def}), one recognizes
that this is exactly the $\kT{1}$ value that yields $\MT2$. This whole point has been first
made in \cite{MAOS}.

As already mentioned in the Introduction, we indicate the $\kT{1}$ choice required by $\MT2$
with $\okT{1}$. We refer to the resulting $k_i$ four-momenta as MAOS momenta, that read
\bea
\label{eq:kiMAOS}
{\bar k}_i^{\pm} ~=~ (\sqrt{m_X^2 + \okT{i}^2 + (\kL{i})^2},\, \okT{i}, \, \kL{i})
~\equiv~ k_i^\maos~,
\eea
with $\okT{2} = \MPT - \okT{1}$, and $\kL{i}$ the solutions of the first two eqs. 
(\ref{eq:on-shell}). Note that the $\pm$ choices are independent for the two decay chains. 
Therefore, MAOS momenta come with a four-fold ambiguity for each event. In the last equality 
of eq. (\ref{eq:kiMAOS}), the MAOS superscript implicitly includes this ambiguity. Henceforth 
in the analysis, when referring to or calculating MAOS-reconstructed observables, it will be 
understood that all of the four solutions are included.

The interest of the MAOS momenta (\ref{eq:kiMAOS}) is in the observation \cite{MAOS}
that $\k{i}^\maos$ are distributed around the true $\k{i}$, even when calculated with
$m_Y$ and $m_X$ values that {\em differ} from the true $Y$ and $X$ masses. (Of course these
masses should still fulfill the inequality $m_Y \ge \MT2(m_X)$ in order to ensure that the
kinematics be in the physical region \cite{ChengHan}. See corresponding discussion in
the Introduction.)
This observation makes MAOS momenta potentially valuable `estimators' of the {\em separate}
invisible momenta $k_i$, in processes of the kind in eq. (\ref{eq:decay}). The question is then 
{\em how well} these estimators actually represent the true, unknown, momenta
of the two invisible particles. In general, this question heavily depends on the process and on
the observable chosen. As detailed in the Introduction, in this paper we confine this question
to the per se interesting case of top polarization and $\ttbar$ spin correlations
in di-leptonic $t \bar t$ decays, eq. (\ref{eq:ttbar_dilep}).
Top-polarization observables (in particular energy ratios and angular distributions) will be 
discussed in sec. \ref{sec:tpol} and spin correlations in sec. \ref{sec:spin_corr}. In the next 
section we will instead address in detail the different MAOS-momenta definitions that are actually 
possible in di-leptonic $t \bar t$ decays.

\subsection[MAOS momenta in di-leptonic $t \bar t$ decays]{\boldmath MAOS momenta in di-leptonic $t \bar t$ decays} \label{sec:MAOS-dilep-tt}

One important aspect of the $\MT2$ variable is the fact that the decay topologies (\ref{eq:decay})
to which it is applicable may consist of one {\em or more} visible particles on each side of the 
decay. $\MT2$ needs as input only the {\em total} visible momentum $p_i$ for decay chain $i$, 
irrespective of how it is composed. This ambiguity allows to construct more than one $\MT2$ variable 
as soon as $p_i$ is the resultant of more than one measurable momentum.

In the case of the di-leptonic $t\bar{t}$ decay (\ref{eq:ttbar_dilep}), to which the rest
of the discussion will be confined, there are three ways of defining the visible particle
system, namely
\begin{enumerate}

\item $b \ell^+$ and $\bar{b} \ell^-$: in this case $m_Y = m_t$ and $m_X = m_\nu = 0$,

\item $\ell^+$ and $\ell^-$: this is a sub-system decay, with parent the $W$ boson, thus
$m_Y = m_W$ and $m_X = 0$,

\item \label{MT2_b}
$b$ and $\bar{b}$: here the $W$ boson should be regarded as the `invisible' particle
on each side of the decay. Then, the missing transverse momentum should be redefined as
$\MPT \to \MPT + \p{\T}^{\ell^+} + \p{\T}^{\ell^-}$. In this case $m_Y = m_t$ and
$m_X = m_W$.

\end{enumerate}
In the first definition, the visible particle masses are $m_{b\ell^+}$ and $m_{\bar{b}\ell^-}$,
whereas they are vanishing (to excellent approximation) in the other definitions. The $\MT2$ variables
calculated from
systems other than the full system are usually referred to as sub-system $\MT2$~\cite{Burns:2008va}.
Depending on the definition of the visible particle system according to the three cases above, we 
indicate the corresponding $\MT2$ variable as $\MT2^{b\ell}$, $\MT2^\ell$, or $\MT2^b$.

The main point is that different $\MT2$ definitions imply different MAOS momenta, that will 
be indicated as $k^{\maos-b\ell}$, $k^{\maos-\ell}$, or $k^{\maos-b}$, respectively.\footnote{%
Obviously, in the case of $\MT2^{b}$, the $k_i$ momenta would correspond to those of the $W$ bosons,
not those of the neutrinos. One can obtain each neutrino momentum by subtracting the known momentum 
of the associated charged lepton. Here $k^{\maos-b}$ denotes the resulting neutrino momentum.
}
One can expect that their performance as estimators of the true invisible momenta be different.
We have made this comparison by generating parton-level Monte Carlo event samples of
$t\bar{t}$ production in $pp$ collisions at 14 TeV c.o.m. energy, using
\textsc{MadGraph 5}~\cite{Alwall:2011uj}. We excluded any kinematic cuts in order to avoid
cut-induced distortions of the phase space. Then, in order to calculate the MAOS momenta, we
chose events with $\MT2$ values equal or smaller than the known $\MT2^{\rm max}$ -- for instance
$\MT2^{b\ell} \leq m_t$ in the case of the full-system $\MT2$.\footnote{%
In our instance, the $\MT2^\max$ value is known in each of the three cases mentioned at the beginning
of sec. \ref{sec:MAOS-dilep-tt}. In cases where it is not known, it can be determined by a functional
fit or a comparison with template distributions, parameterized in terms of $m_Y$. In general, the
$\MT2$ distribution displays a tail above $\MT2^\max$, due to finite decay width as well as
unreliable evaluations.}
Explicitly imposing this condition allows to minimize the number of events where the $\MT2$ algorithm 
fails to find the correct minimum. (This occurs more frequently in events very close to the end-point,
as they also get close to the boundary of the physical region \cite{ChengHan}. In our simulation,
the fraction of such events is very small anyway, about $0.4\%$.)
We note in passing that an $\MT2$ upper cut may also reveal itself useful for analyzing 
detector-level data where particle identification and detector-resolution effects occasionally 
make the $\MT2$ calculation badly fail.

\begin{figure}[tb!]
\begin{center}
  \includegraphics[width=0.48\textwidth]{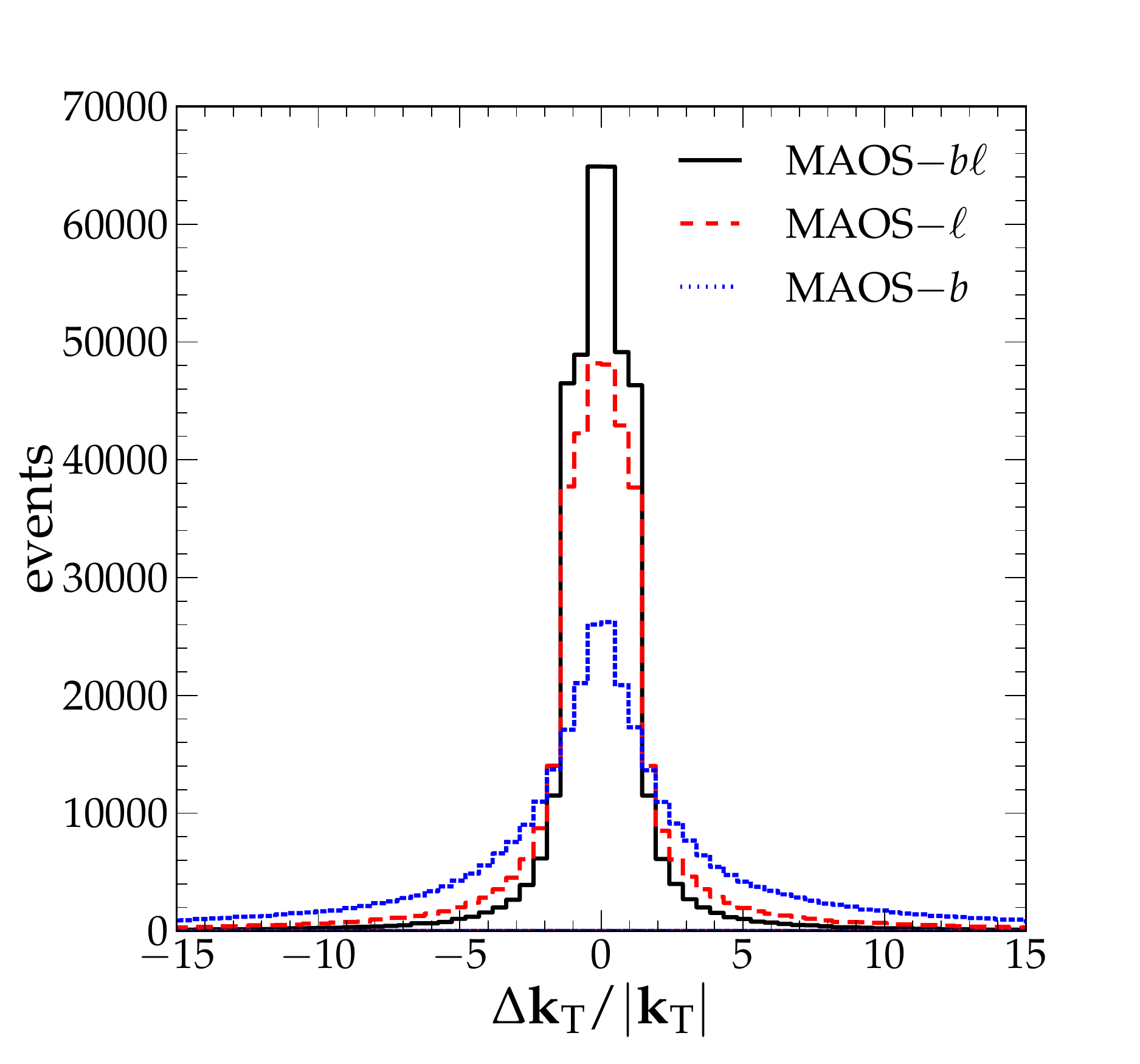}
  \includegraphics[width=0.48\textwidth]{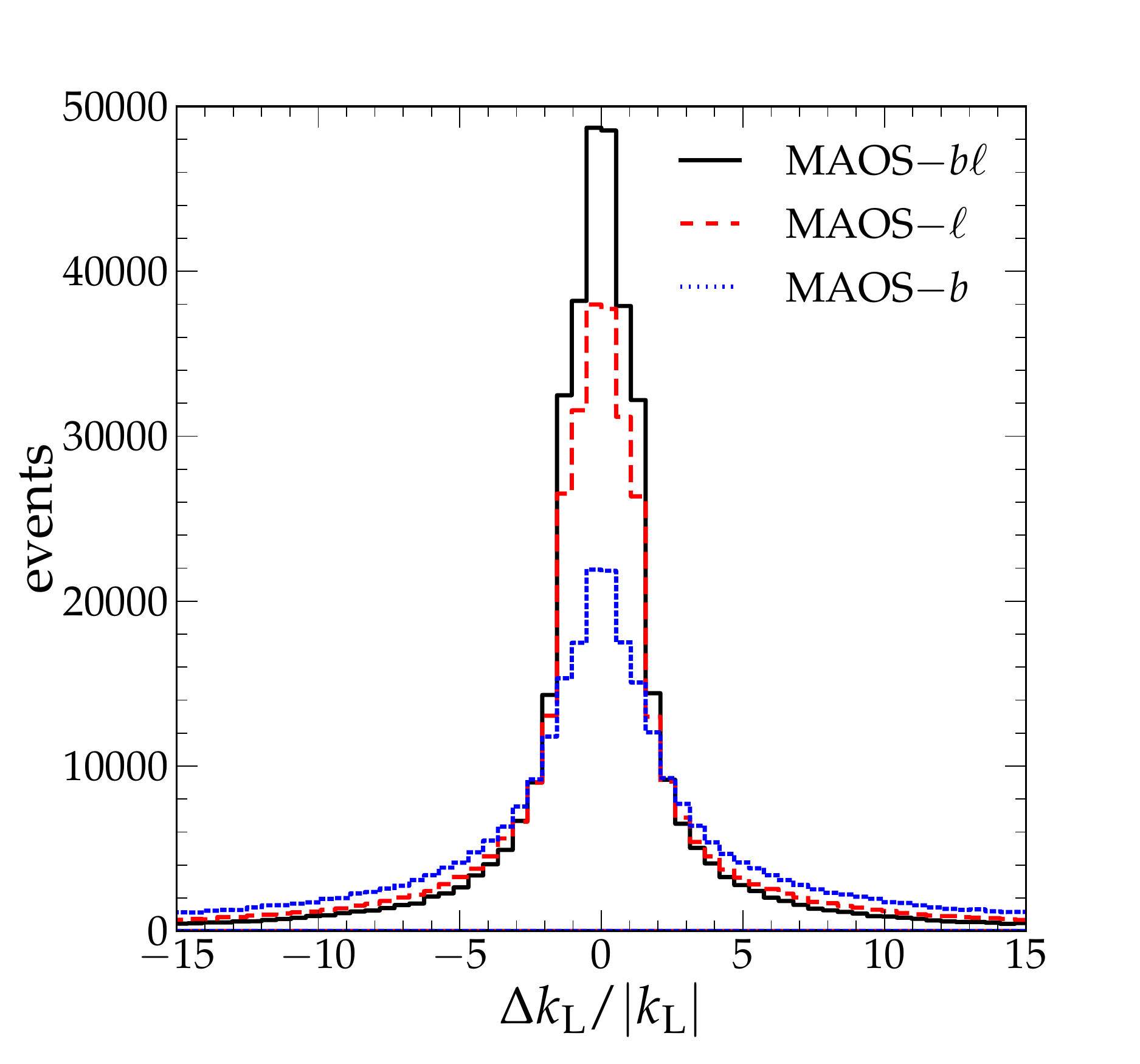}
\end{center}
  \caption{Distributions of
  $\Delta \k{\T} / |\k{\T}| \equiv (\k{\T}^\maos - \k{\T}^\true) / |\k{\T}^\true |$ (left panel)
  and
  $\Delta k_{\rm L} / |k_{\rm L}| \equiv (k_{\rm L}^\maos - k_{\rm L}^\true) / |k_{\rm L}^\true|$
  (right panel). $k$ is the neutrino momentum in SM $t\bar{t}$ production (LHC, 14 TeV) followed
  by a decay to $b \ell \nu$ on both sides. The $x$-axis variable in the left panel stands 
  for either of $\Delta k_x$ or $\Delta k_y$.}
\label{fig:delta_k}
\end{figure}
Fig.~\ref{fig:delta_k} displays the distributions of $\Delta \k{\T} / |\k{\T}| \equiv
(\k{\T}^\maos - \k{\T}^\true) / |\k{\T}^\true |$ for the different MAOS momenta, showing that
$k^{\maos-b\ell}$ matches best the true neutrino momenta. From the distributions one sees that
the vast majority of the MAOS-estimated invisible momenta differ from the true momenta by less
than a factor of two in either of the transverse and the longitudinal directions.
The $k^{\maos-\ell}$ performs somewhat worse than $k^{\maos-b\ell}$, whereas $k^{\maos-b}$ is 
not comparable to the others. Henceforth we will thus focus on $k^{\maos-b\ell}$ and $k^{\maos-\ell}$.

\begin{wrapfigure}{r}{0.5\textwidth}
  \vspace{0.2cm}
  \hspace{5pt}
  \includegraphics[width=0.48\textwidth]{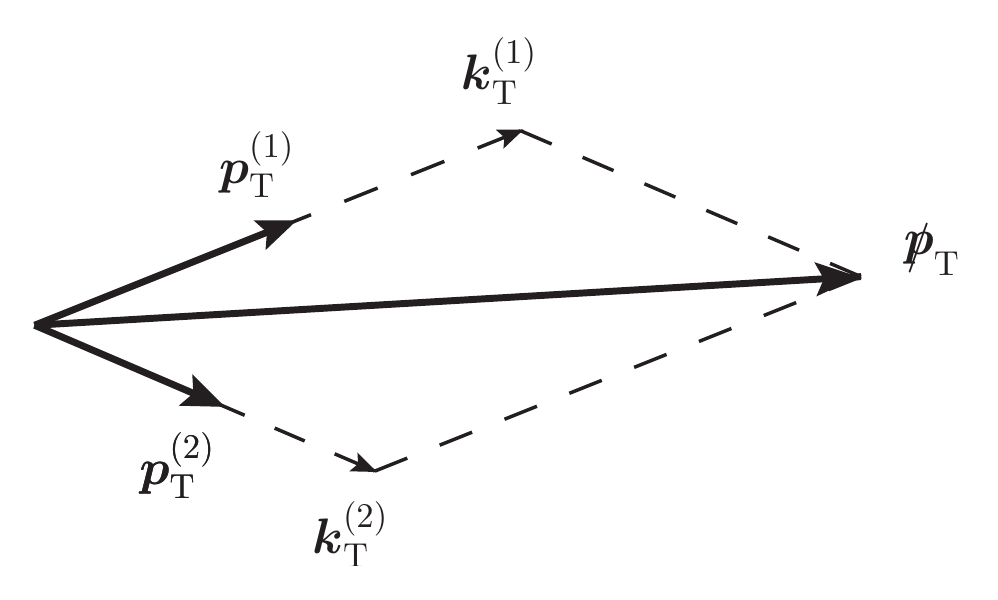}
    \caption{Kinematic configuration for the trivial-zero MAOS solution. See text for details.}
  \label{fig:pT12cone}
\end{wrapfigure}

The difference of efficiency between $k^{\maos-b\ell}$ and $k^{\maos-\ell}$ is due to two main
reasons. The first one is the trivial $\MT2$ zero when all the visible and invisible particles
are massless, as is the case for $\MT2^{\ell}$~\cite{Lester:2011nj}. The trivial-zero solution occurs
when the missing transverse momentum $\MPT$ lies inside of the smaller of the two cones enclosed
between the visible-particle momenta $\p{\T}^{(1)}$ and $\p{\T}^{(2)}$ (see Fig. \ref{fig:pT12cone}).
In such case, the $\MT2$ value is attained for a momentum configuration where $\k{\T}^{(i)}$
is proportional to its visible partner momentum $\p{\T}^{(i)}$, thus making the transverse masses in
eq.~(\ref{eq:MT2def}) all vanish.

\begin{figure}[tb!]
  \begin{center}
    {\includegraphics[width=0.48\textwidth]{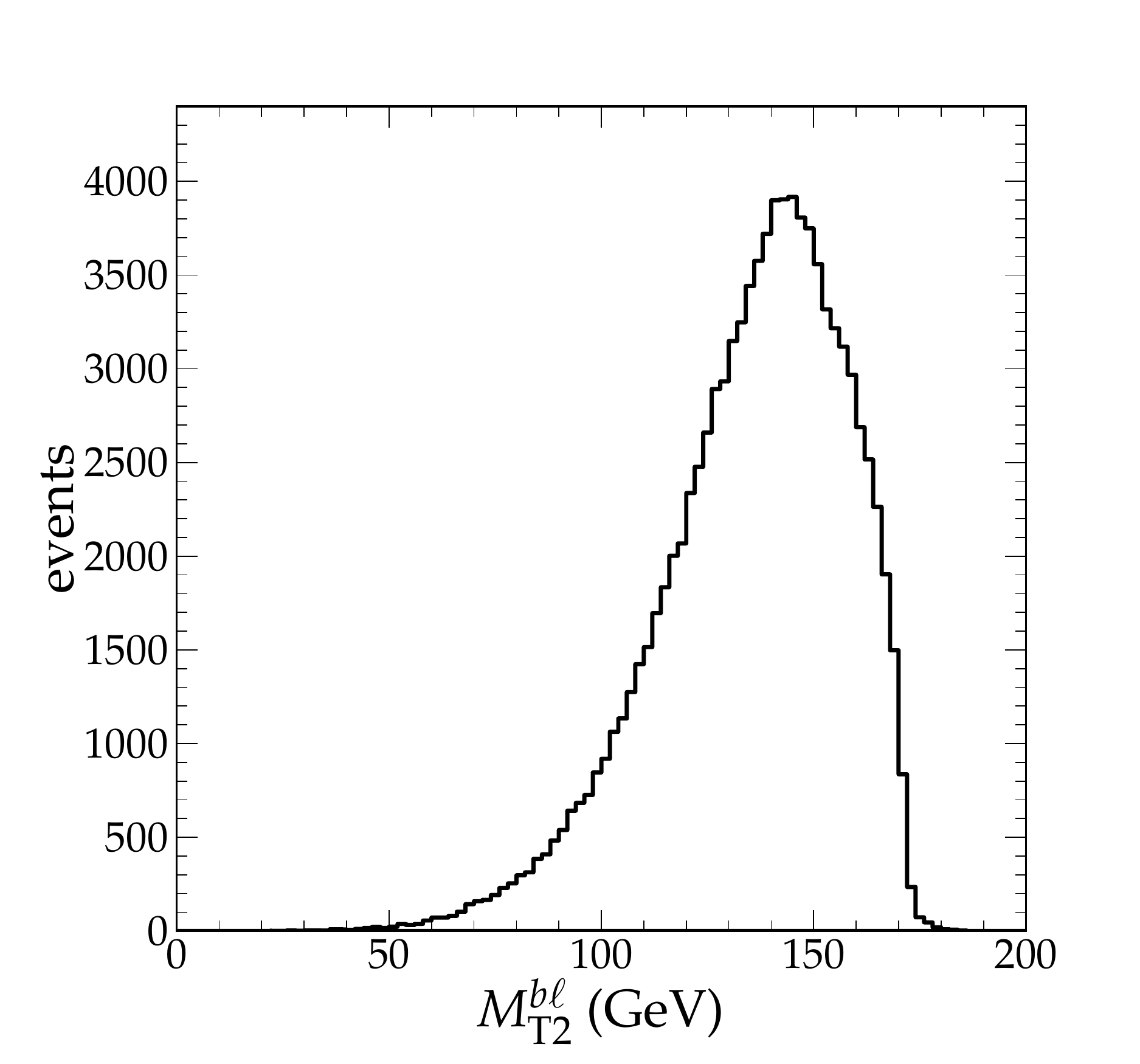}
      \includegraphics[width=0.48\textwidth]{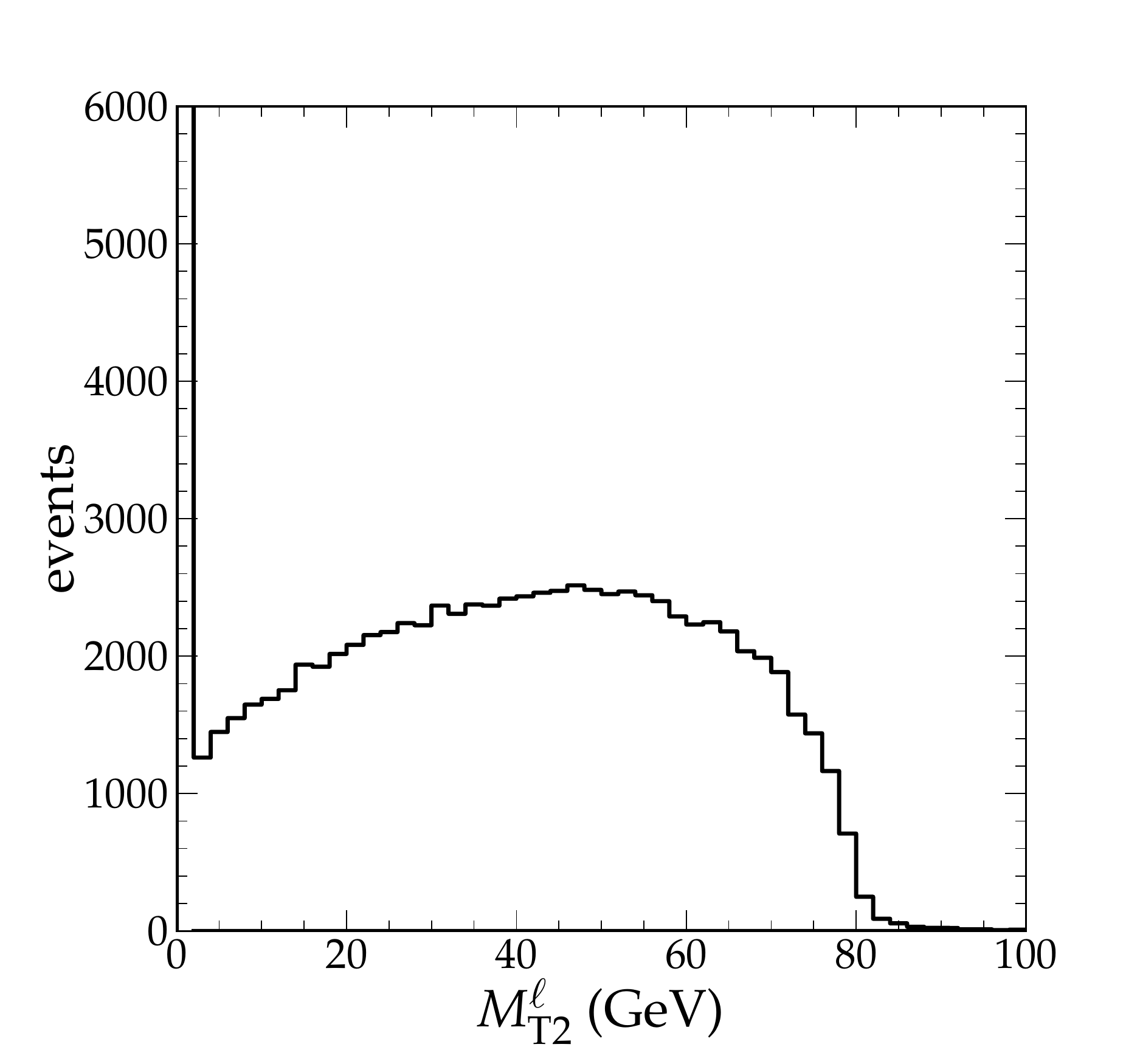}}
    {\includegraphics[width=0.48\textwidth]{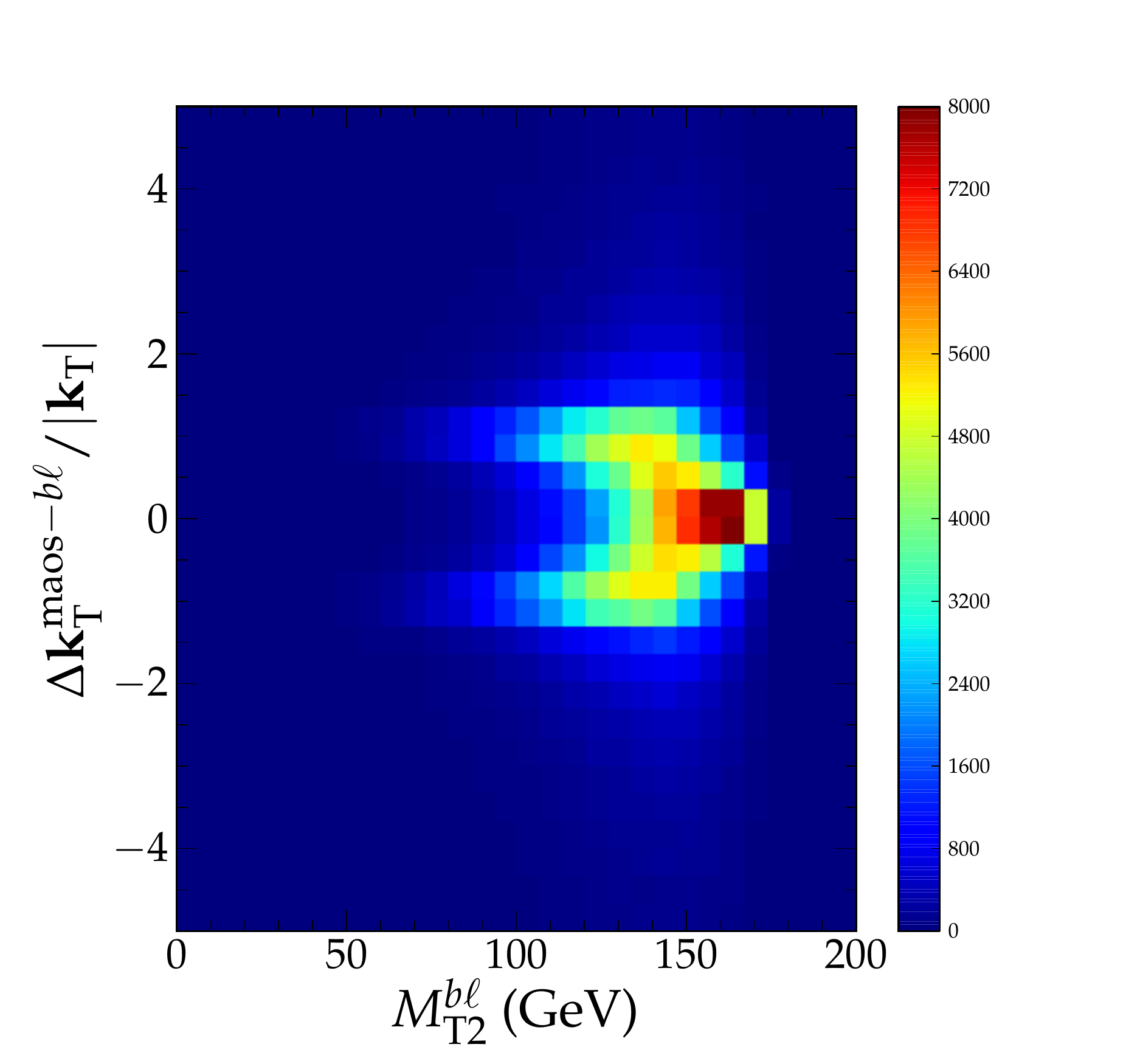}
      \includegraphics[width=0.48\textwidth]{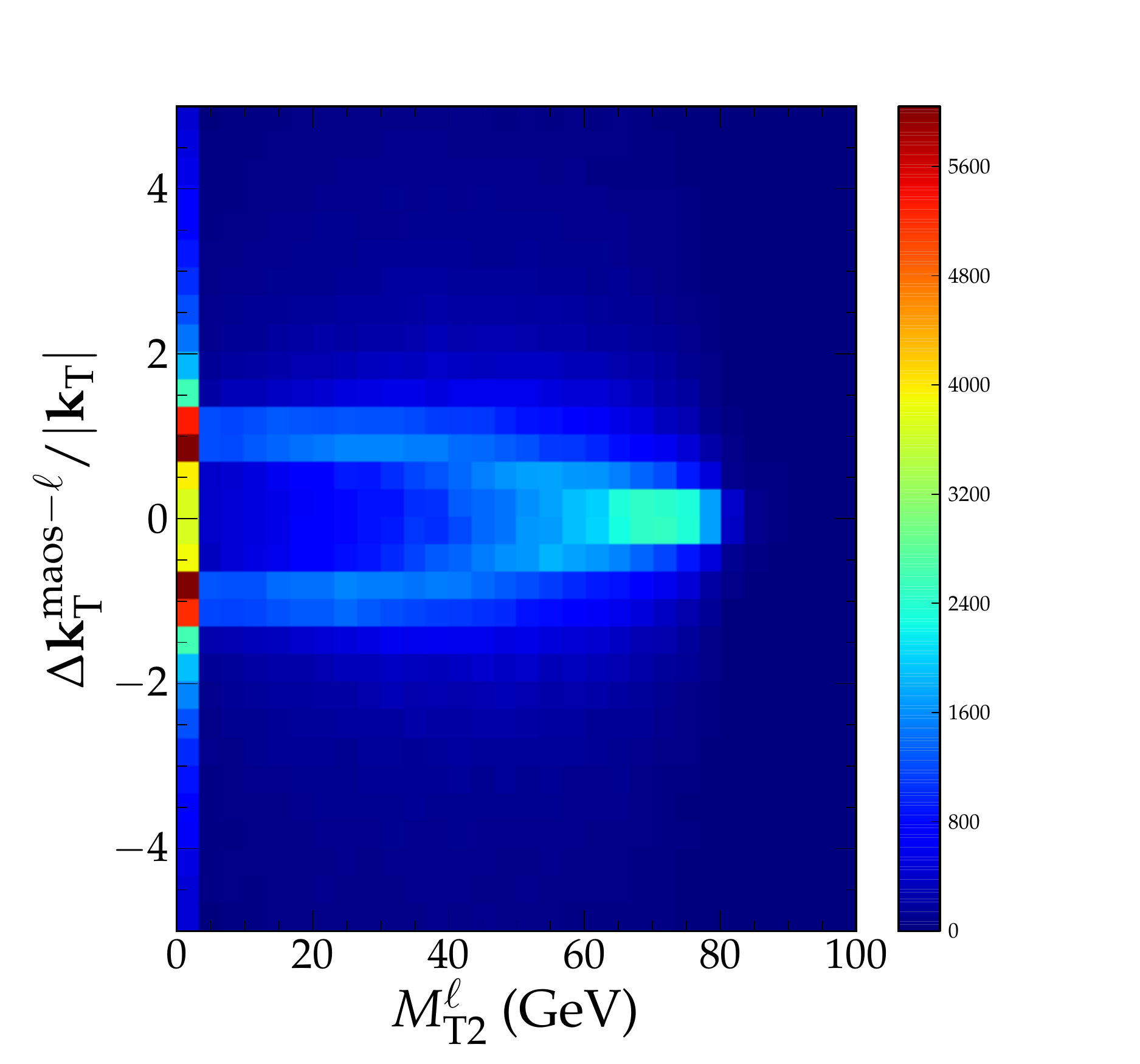}}
  \end{center}
  \caption{Distributions of (upper frame) $\MT2$ and (lower frame) the 
  correlation between $\MT2$ and $\Delta \k{\T}$ for (left panels) 
  $k^{\maos-b\ell}$ and (right panels) $k^{\maos-\ell}$.}
  \label{fig:delta_kt_mt2}
\end{figure}

\indent The right panels of Fig.~\ref{fig:delta_kt_mt2} show indeed a tower
of events in the lowest $\MT2^{\ell}$ bin. Application of the MAOS method to real situations requires
a suitable $\MT2$ cut, excluding events with too small $\MT2$ values. In fact, as stated in
\cite{MAOS}, the MAOS algorithm performs best in events with $\MT2$ values closer to the
endpoint. Therefore, the trivial-zero solution does not set a fatal limitation to the MAOS method as
long as a reasonable $\MT2$ lower cut is imposed.

The second reason for the different efficiency between $k^{\maos-b\ell}$ and $k^{\maos-\ell}$
is the number of kinematic configurations close to the $\MT2^{\rm max}$ value -- as we just said, 
the region where the MAOS algorithm performs best.
For instance, in the case of $\MT2^{b\ell}$ the visible-particle systems consist
of one $b$ quark and one charged lepton. By construction, $\MT2^{b\ell}$ depends only on the sum
of their momenta, irrespective of the individual momentum magnitudes. This freedom implies that the
same $\MT2$ value can be attained with different choices of these individual momenta, and the number
of these choices is higher for higher $\MT2$ values. Hence (many) more kinematical configurations
close to the $\MT2$ endpoint are possible for $\MT2^{b\ell}$ than for $\MT2^{\ell}$ and this 
explains why, close to the endpoint, the $\MT2^{b\ell}$ peak is much sharper than the 
$\MT2^{\ell}$ one (cf. left vs. right panels of Fig. \ref{fig:delta_kt_mt2}).\footnote{%
It is worth noting that in reality also momenta flowing upstream with respect to the decay
process of interest -- e.g. initial-state radiation -- can play some role to make $\MT2$
maximal~\cite{Gripaios_UTM,Burns:2008va,Mahbubani,Cho:2009wh}. We confine our discussion 
to the case of vanishing upstream transverse momentum for the sake of simplicity.}

While the first reason (trivial-zero solution) does not occur in cases where the visible and invisible
particles are massive, the second reason holds in general: the larger the number of ways to compose
individual momenta to obtain the same total visible momentum (the only input needed for $\MT2$), the
larger the number of events close to the $\MT2$ endpoint, where MAOS performs best. From this one can
deduce that, in cascade decays involving several steps, the best-performing MAOS momenta are those
constructed from the full-system $\MT2$.

A final remark on the possible role of initial-state QCD radiation (ISR) is in order. The latter is 
known \cite{Gripaios_UTM,Mahbubani} to potentially affect the $\MT2$ near-endpoint region, that, as 
we have been arguing, is the region where the MAOS algorithm performs best. We have checked the 
negligibility of this effect by studying the distributions in Fig. \ref{fig:delta_k} for events
with an additional ISR jet, and found no appreciable differences.\footnote{An argument in support of 
this statement is provided by fig. 12 of the second ref. in \cite{Gripaios_UTM}.}

Using $k^{\maos-b\ell}$ we have now a systematic way of estimating the two neutrino momenta, and
of thereby reconstructing the $t$ and $\bar t$ rest frames. The latter can be used to evaluate
top-polarization or $\ttbar$ spin-correlation observables, to be studied in the next sections.

\section{Top polarization}\label{sec:tpol}

Top decay products obey angular distributions that are correlated with the parent-top spin. This well
known fact is, among quarks, a unique property of the top, and it is due to its large mass. The latter 
is responsible for the top-quark's small lifetime, $\sim 1/(G_F m_t^3)$, much shorter than the time, 
$\sim m_t/\Lambda_{\rm QCD}^2$, needed by QCD interactions to decorrelate the production-time spin 
configuration \cite{Bigi}.

At hadron colliders, top quarks are produced predominantly as $t \bar t$ pairs by QCD processes,
which a priori cause left and right polarizations to weigh equally in an event set. However, in
new-physics scenarios involving chiral couplings, top-quark polarizations may be produced in
unequal weights. Accurately measuring the top polarization is therefore considered as an important
clue for physics beyond the SM.

One can construct at least two different classes of observables measuring top polarization: energy
ratios and the angular distributions of top decay products. We discuss them in turn in the next
two subsections.

\subsection{Top polarization from ratios of daughter to parent particle energies} \label{sec:tpol_Eratio}

One way to test top polarization is via the energy spectra of the top decay products, that may namely
be peaked towards softer or harder values depending on the top being left- or right-handed. We will
focus here on top production followed by a leptonically-decaying $W$, $t \rightarrow b W (\rightarrow
\ell^+ \nu)$. In Refs.~\cite{Czarnecki:1990pe,Schmidt:1992et} it was pointed out that the chirality of 
the top quark is correlated with the ratio
\be
  x_\ell = \frac{2 E_{\ell^+}}{E_t}
  \label{eq:x_l}
\ee
between the charged-lepton energy and that of its parent top quark. Specifically, charged leptons 
produced by right-handed top quarks tend to be more energetic than those produced by 
left-handed top quarks, the difference increasing with the top energy. This conclusion follows from
the fact that the $b$-quark is (to very good approximation) always produced left-handed and the $W$ 
predominantly longitudinal~\cite{Czarnecki:1990pe,Schmidt:1992et}.

\begin{figure}[tb!]
  \begin{center}
    \includegraphics[width=0.48\textwidth]{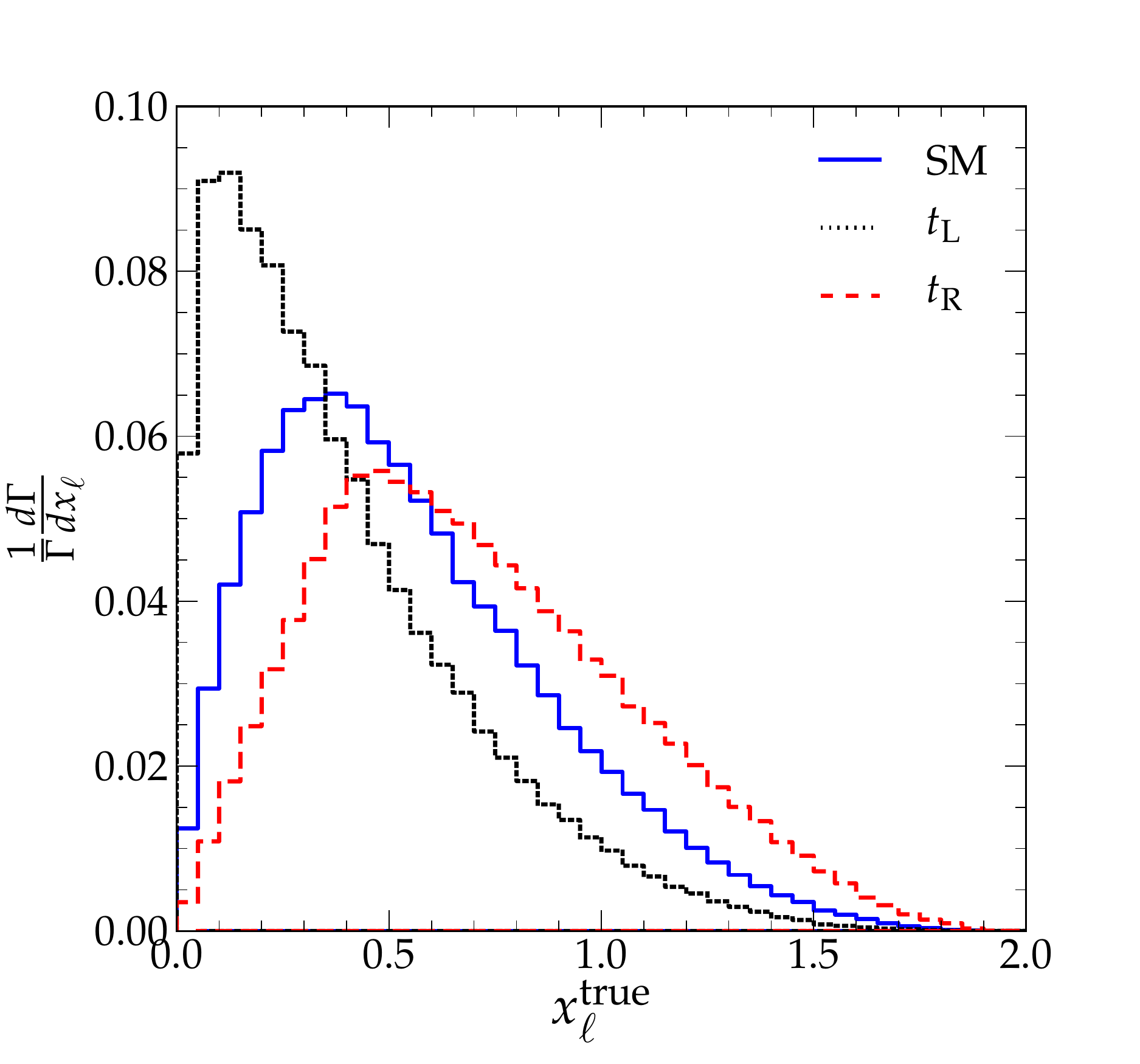}
    \includegraphics[width=0.48\textwidth]{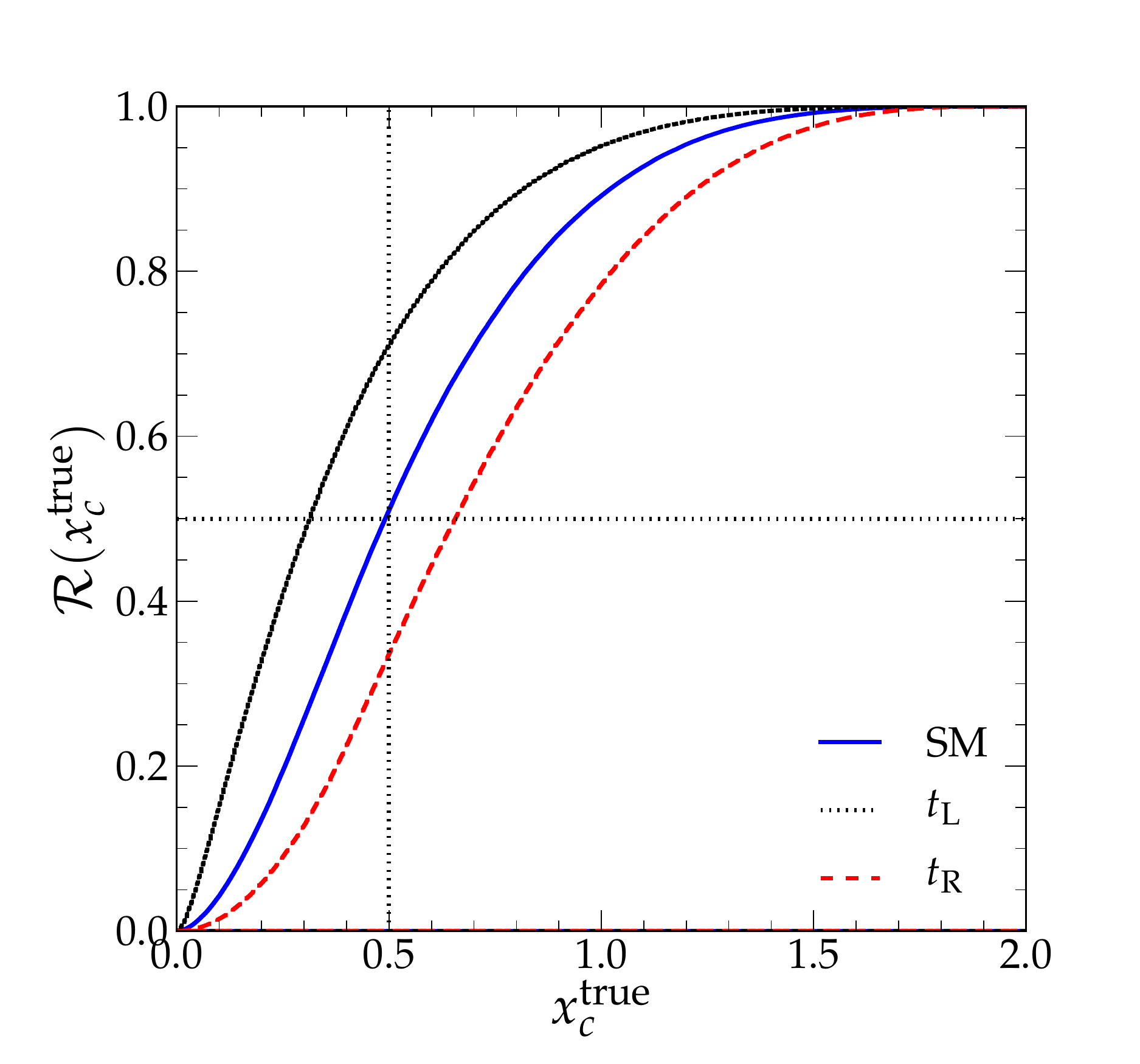}
  \end{center}
  \caption{(Left panel) differential distribution of the energy ratio in eq. (\ref{eq:x_l})
  and (right panel) corresponding cumulative distribution. The three distributions in either plot
  refer to purely left-handed, or purely right-handed, or QCD-produced tops (see legend).
  The antitop branches of each event are also included.}
  \label{fig:x_l_true}
\end{figure}

This feature can easily be checked quantitatively in $pp$ collisions at 14 TeV by generating Monte 
Carlo events for purely left-handed, purely right-handed, or SM-produced $\ttbar$ pairs.\footnote{%
Our Monte Carlo results are, as elsewhere in the paper, generated with 
\textsc{MadGraph 5}~\cite{Alwall:2011uj}. In particular, the purely left- and right-handed $\ttbar$ 
cases are simulated via a toy model containing a new vector with chiral couplings to quarks, 
implemented in \textsc{MadGraph} via \textsc{FeynRules} \cite{FeynRules}.}
The resulting $d \Gamma(t \rightarrow b \ell^+ \nu) / d x_\ell$ distribution at parton level is shown in 
the left panel of Fig.~\ref{fig:x_l_true}.
The difference between the $t_L$ and $t_R$ cases can be better appreciated via the integral of 
the differential distribution up to a given $x_\ell = x_c$ value:
\be
  \mathcal{R}(x_c) \equiv \frac{1}{\Gamma} \int_0^{x_c}
  \frac{d \Gamma}{d x_\ell} d x_\ell~.
  \label{eq:x_c}
\ee
Qualitatively, this cumulative distribution estimates how early the differential distribution 
approaches the peak. The cumulative distributions of the histograms in Fig.~\ref{fig:x_l_true}
(left panel) are shown in the right panel of the same figure.

This strategy can be applied to the extent that the top energy can be reconstructed. For example,
in $t\bar t$ decays where one top decays semi-leptonically and the other hadronically, the momentum 
of the semi-leptonically decaying top can be reconstructed by using the on-shell relations
\begin{align}
  \left ( p^b + p^\ell + k^\nu \right )^2 &= m_t^2 \quad
  \mbox{or} \quad
  \left ( p^\ell + k^\nu \right )^2 = m_W^2 , \nonumber\\
  (k^\nu)^2 &= 0, \quad \k{\T}^\nu = \MPT~.
  \label{eq:on_shell_semi}
\end{align}
Hence $x_\ell$ is calculable from eq.~(\ref{eq:on_shell_semi}) up to a discrete degeneracy.

More generally, however, the $t \bar t$ pair may be the result of a longer decay chain, involving
further undetected particles than just a neutrino -- for example supersymmetric $\tilde t \, \tilde t$
production would lead to $\ttbar$ plus two additional neutralinos. In this case $E_t$ cannot be 
reconstructed directly and it is meaningful to search for `proxies' of the variable in 
eq.~(\ref{eq:x_l}), that do not involve $E_t$. This issue has been recently explored 
in~\cite{Berger:2012an}.\footnote{%
\label{foot:Perez}%
Another instance in which one can construct top-polarization observables without the need to 
reconstruct the top rest frame is when tops are highly boosted~\cite{Perez,Shelton:2008nq,Nojiri},
as is the case if they are produced from accordingly massive new physics.}

\begin{figure}[tb!]
  \begin{center}
    {
    \includegraphics[width=0.48\textwidth]{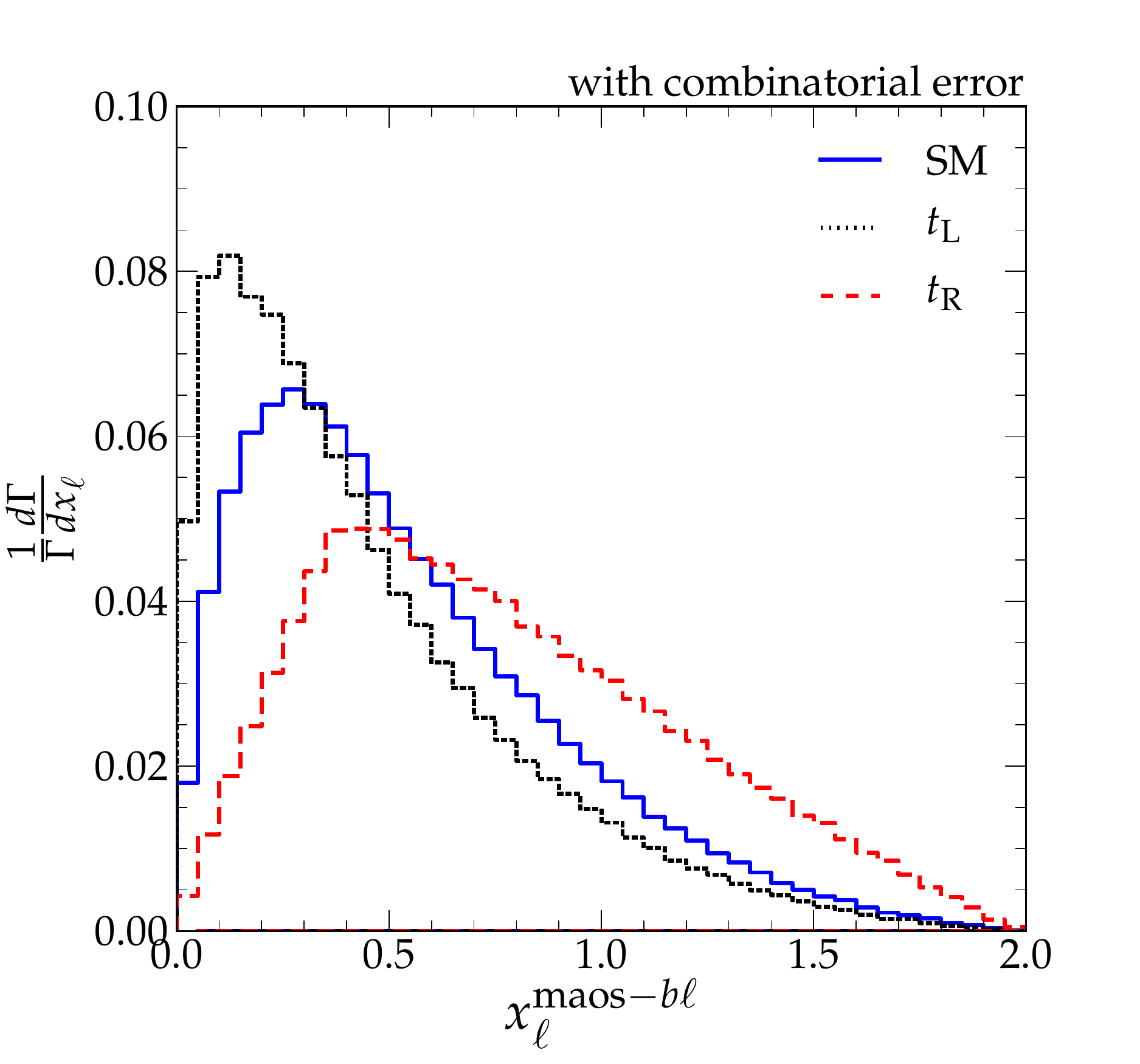}
    \includegraphics[width=0.48\textwidth]{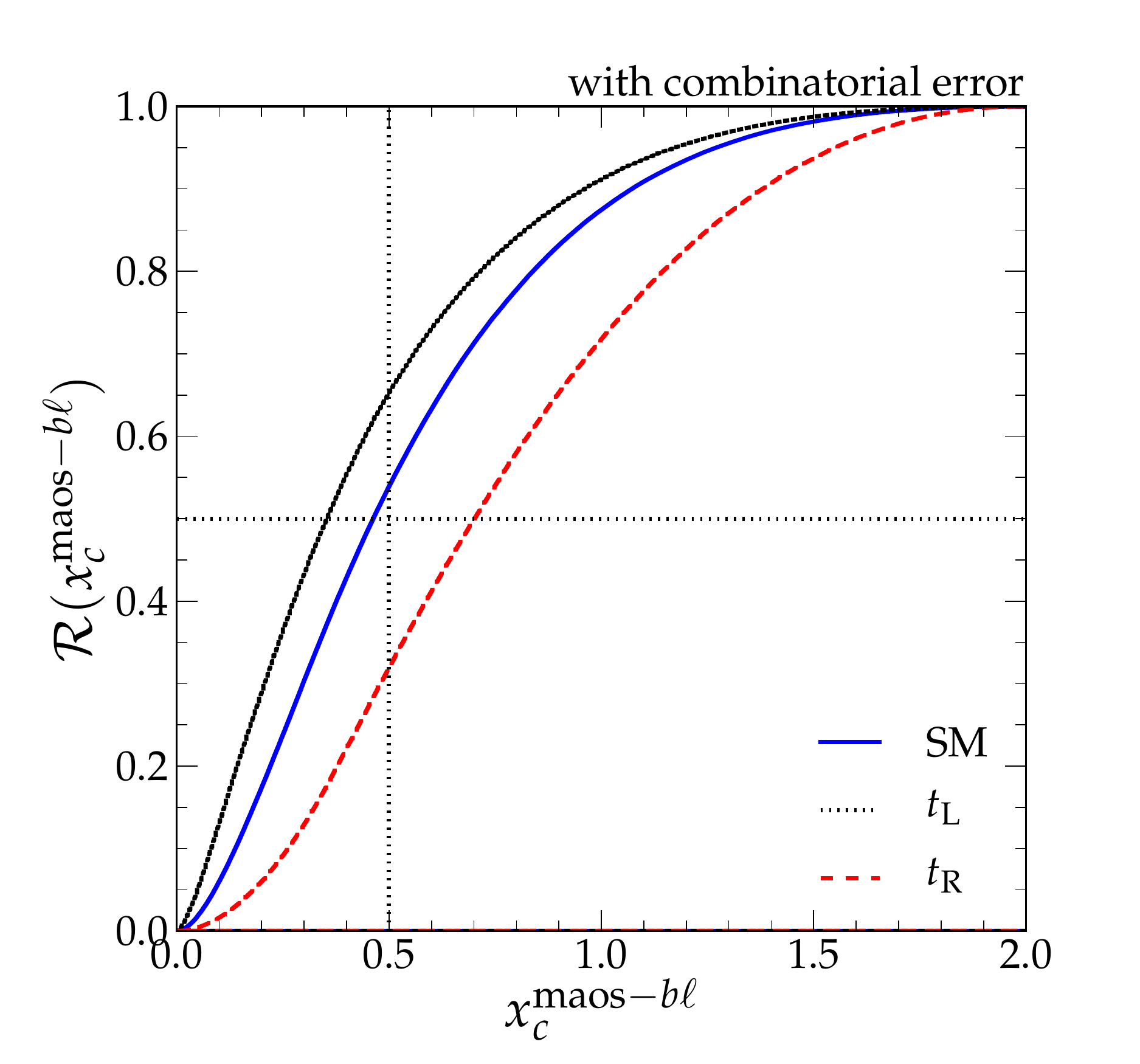}
    }
    {
    \includegraphics[width=0.48\textwidth]{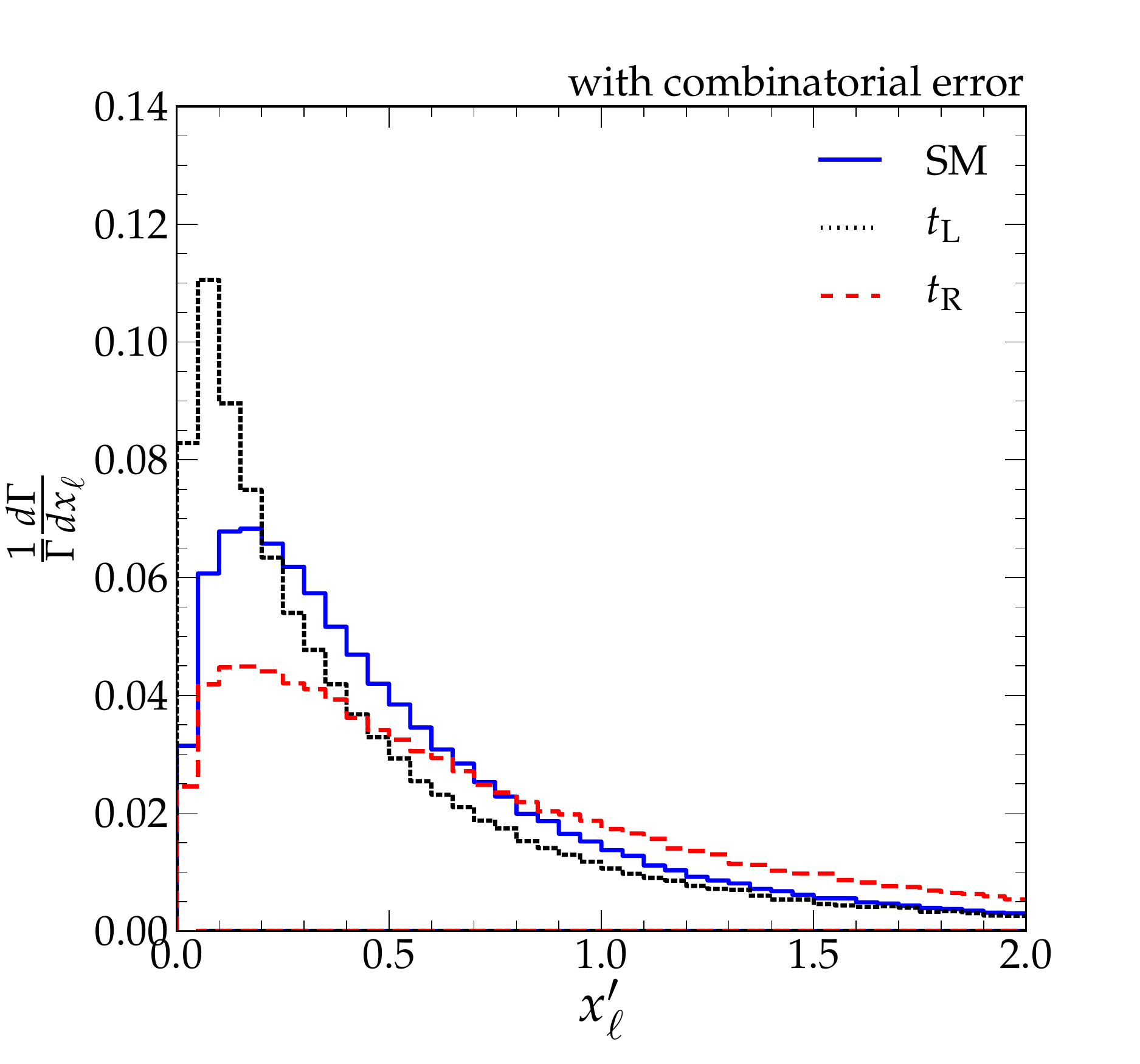}
    \includegraphics[width=0.48\textwidth]{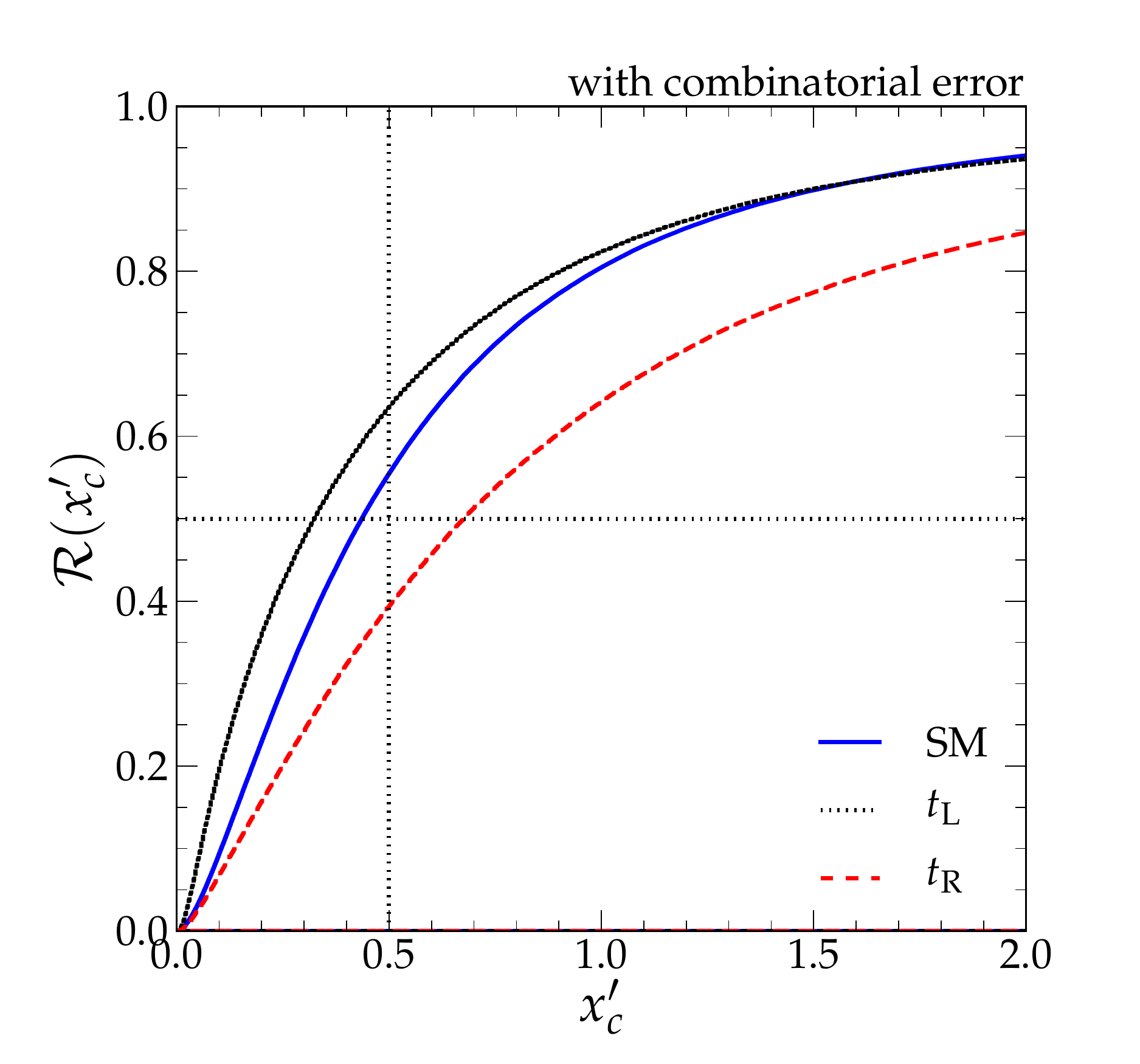}
    }
  \end{center}
  \caption{Same as Fig. \ref{fig:x_l_true}, but with the $x_\ell$ variable evaluated with
  the MAOS method, as in eq. (\ref{eq:x_l_maos}) (upper frames), or using the primed
  variable in eq. (\ref{eq:x_l'}) \cite{Berger:2012an} (lower frames).}
  \label{fig:x_l_maos}
\end{figure}
In particular, the authors of \cite{Berger:2012an} consider the case where one of the tops decays
as $t \rightarrow b \ell^+ \nu$ and the other as $\bar t \rightarrow \bar b q \bar q^\prime$. The 
energy of the semileptonically-decaying top, $E_t$, is estimated, event by event, via the energy 
of the other (anti-)top, $E_{\bar{t}}$, that at least in principle is measurable. They thus 
introduce the modified energy-ratio variable
\be
  x_\ell^\prime = \frac{2 E_{\ell^+}}{E_{\bar{t}}}
  = \frac{2 E_{\ell^+}}{E_{\bar{b}} + E_{q} + E_{\bar q^\prime}}.
  \label{eq:x_l'}
\ee

The above strategy cannot (or at least is not designed to) be applied in $t \bar t$ events where 
both $W$ decay to leptons, because the two undetected neutrinos in the final state challenge the
reconstruction of both $E_t$ and $E_{\bar t}$. On the other hand, since the $t \bar t$ 
decay topology is suitable for the construction of $\MT2$, the parent-particles' energies can actually 
be estimated using the MAOS method discussed in the previous section. We accordingly define
\be
  x_\ell^\maos = \frac{2 E_{\ell^+}}{E_t^\maos} =
  \frac{2 E_{\ell^+}}{E_b + E_{\ell^+} + E_\nu^\maos},
  \label{eq:x_l_maos}
\ee
where $E_\nu^\maos = |\k{}^{\maos-b\ell}|$. Here we choose the MAOS four-momentum estimated from the
full-system $\MT2$, $k^{\maos-b\ell}$ (cf. discussion in sec. \ref{sec:MAOS-dilep-tt}).\footnote{%
Note that, event by event, the $m_t$ and $m_W$ masses that enter the $\MT2$ calculation float 
according to their finite widths. This effect is taken into account in all of our numerics.}
The differential and respectively cumulative (the analogue of eq.~(\ref{eq:x_c})) distributions of 
$x_\ell^\maos$ are shown in the upper frames of Fig.~\ref{fig:x_l_maos}. As a comparison, the
corresponding distributions for the case of the $x_\ell^\prime$ variable in eq.~(\ref{eq:x_l'}) are 
shown in the lower frames of Fig.~\ref{fig:x_l_maos}.

It should be noted that application of the $x_\ell$ variable to di-leptonic $\ttbar$ decays comes, 
by construction, with an additional uncertainty, namely the two-fold combinatorial ambiguity of 
correctly assigning the 2 $b$-jets (that are not flavor-tagged, i.e. their charge is not determined 
in general) + $2 \ell$ final state to the two decay chains. We address this ambiguity using the method 
in \cite{Choi_MT2comb}. 
Energy-ratio variables, such as those considered in this section, turn out to be rather robust 
with respect to the combinatorial error: distributions where this error is taken
into account barely differ with respect to those with final states always paired correctly. Hence
in this section we only show distributions where this ambiguity is included.\footnote{We will return 
to this issue in much more detail in secs. \ref{sec:tpol_ang} and \ref{sec:spin_corr}, where its 
interplay with the MAOS method and the cuts leads to more insights on our method.}

The following comments on Fig. \ref{fig:x_l_maos} are in order.
\begin{enumerate}

\item Since the computation of the MAOS momentum preserves energy-momentum conservation, $E_\ell$
is always smaller than $E_t^\maos$, hence the $x_\ell^\maos$ distribution has a definite cutoff at
2, like the $x_\ell^{\rm true}$ distribution constructed with the true $E_t$, and shown in
Fig.~\ref{fig:x_l_true}. Note that, on the other hand, the $x_\ell^\prime$ differential distribution
does not fulfill the same cutoff requirement, as confirmed by the lower plots of Fig.~\ref{fig:x_l_maos}.

\item In the cumulative $x_c^\maos$ distribution, the unpolarized case (the SM one) lies neatly between
the purely $t_L$ and the purely $t_R$ cases, in close resemblance to the true distribution.
Again, this is largely consequence of the fact that the MAOS distributions fulfill the cutoff 
constraint mentioned in item 1.

\item From the previous items, one concludes that the distribution constructed with the MAOS method
is fairly close to the true distribution already at the differential level. Thus this method allows
to test top polarization via the $x_\ell$ variable, even in the di-leptonic $t \bar t$ decay channel.

\end{enumerate}

\begin{figure}[tb!]
  \begin{center}
    {
    \includegraphics[width=0.48\textwidth]{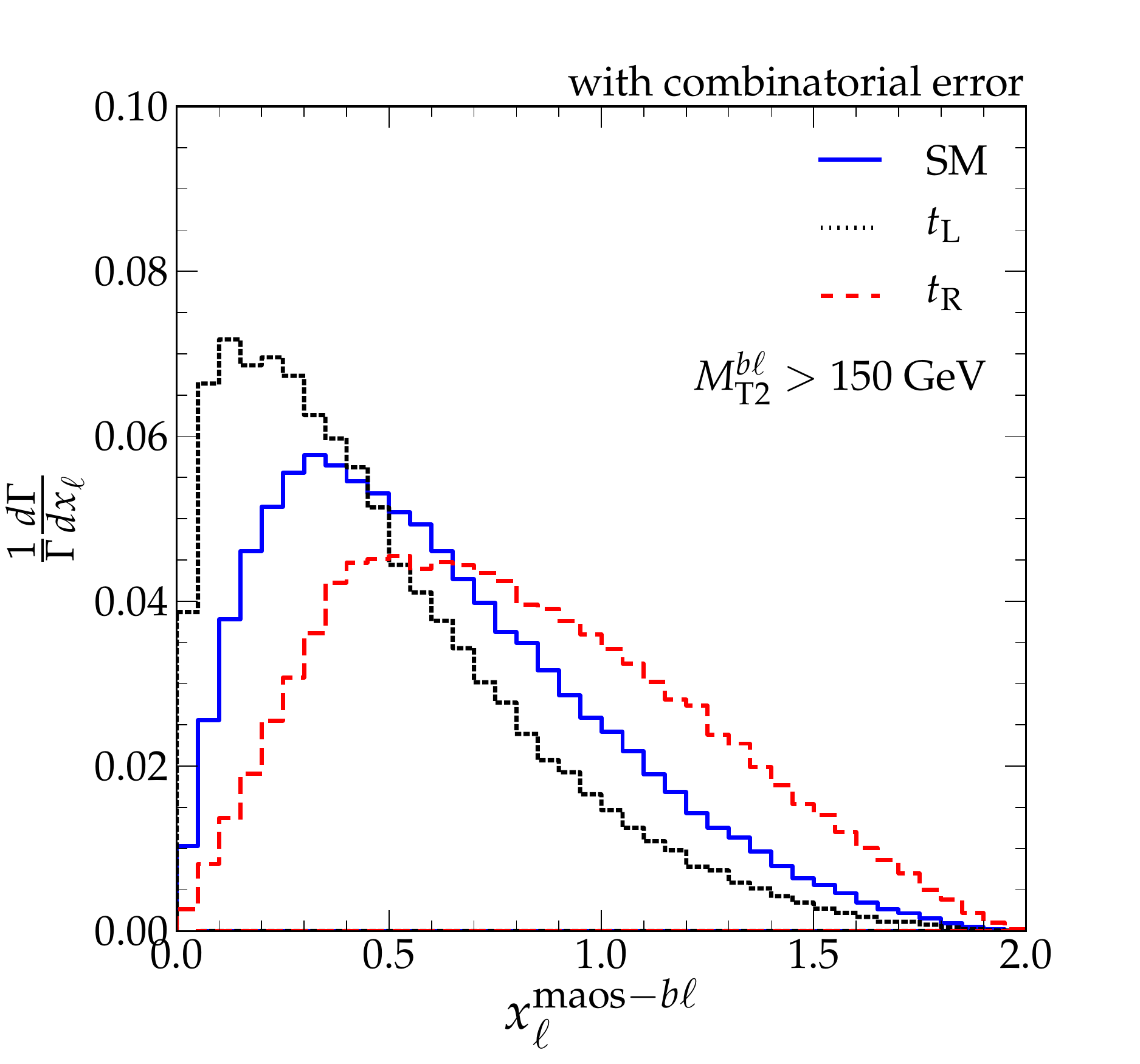}
    \includegraphics[width=0.48\textwidth]{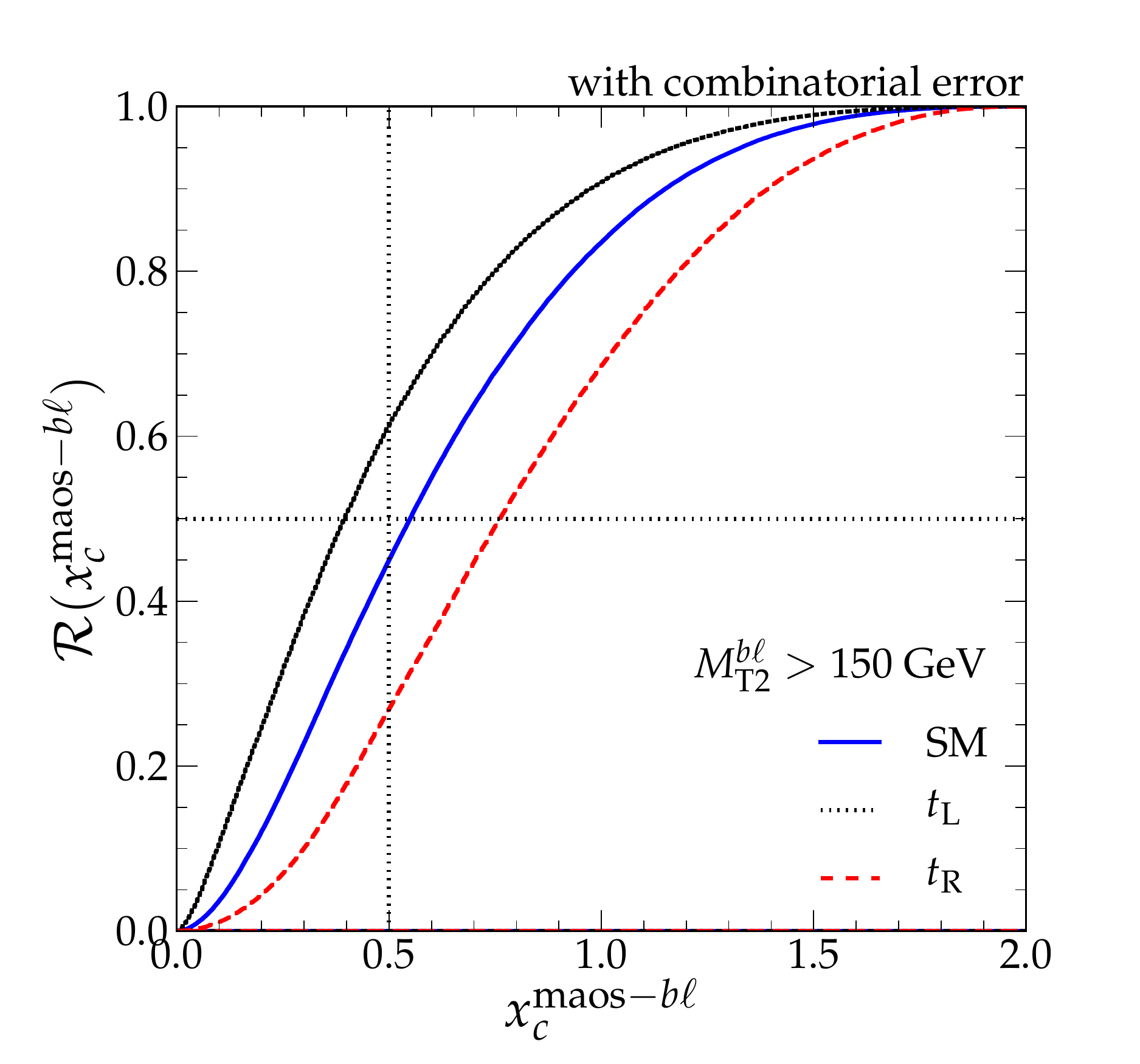}
    }
  \end{center}
  \caption{Same as the upper panels of Fig. \ref{fig:x_l_maos}, but including a lower cut on
  $\MT2$.}
  \label{fig:x_l_maos_bl_cut}
\end{figure}
A further virtue of the MAOS method is that the accuracy of the approximation is under the control
of the $\MT2$ cut. As mentioned, the MAOS momenta become closer to the true momenta for events
with $\MT2$ approaching $\MT2^{\rm max}$, as shown in the lower left frame of
Fig.~\ref{fig:delta_kt_mt2}. Therefore, by imposing a suitable $\MT2$ cut, the accuracy of the MAOS
momentum can be increased at the expense of statistics. Fig.~\ref{fig:x_l_maos_bl_cut} shows the
same distributions as in the upper frames of Fig.~\ref{fig:x_l_maos} but with the inclusion of an
$\MT2^{b\ell} > 150$ GeV cut. Note that, because the $\MT2$ distribution has a peak structure close
to the endpoint (see left panels of Fig. \ref{fig:delta_kt_mt2}), the number of events not passing
this cut is (only) half of the total dataset. By comparing Fig. \ref{fig:x_l_maos_bl_cut} with
Figs. \ref{fig:x_l_maos} and \ref{fig:x_l_true}, one can see that the cumulative MAOS distribution
is close to the true distribution, and that the distribution with the $\MT2$ cut gets even closer 
to it. In fact, the inclusion of the $\MT2$ cut is above all intended to check explicitly that it 
does not introduce distortions in the overall distributions. This would occur if the cut selected 
kinematic configurations more populated e.g. by $t_R$ than by $t_L$, so as to introduce cut-induced 
asymmetries.

\subsection{Top polarization from angular variables} \label{sec:tpol_ang}

The MAOS method allows full reconstruction of the parent-particle's momentum. This permits to test 
the most direct of top-polarization observables, the angular distribution of top decay products. 
Among the latter, charged leptons have the double advantage of a `maximal' spin-analyzing power 
\cite{JezabekKuhn} and of being especially clean objects for experiments. At tree level, the 
charged-lepton distribution in top-quark decays can be written as (see e.g. \cite{MahlonParke1})
\be
  \frac{1}{\Gamma} \frac{d \Gamma}{d \cos\theta} =
  \frac{1 + \alpha \cos\theta}{2},
  \label{eq:dGamma_dcostheta}
\ee
where $\theta$ denotes the angle between the decaying-particle spin-quantization axis and the
direction of the charged lepton, viewed in the decaying-particle's rest frame. The coefficient
$\alpha$ denotes the mentioned charged-lepton spin-analyzing power, equalling $+1~(-1)$ for spin-up 
(spin-down) tops or spin-down (spin-up) antitops. Angular distributions from decay products other
than charged leptons obey relations entirely analogous to eq. (\ref{eq:dGamma_dcostheta}), but 
for a different spin-analyzing power $|\alpha| \leq 1$. 

\begin{wrapfigure}{l}{0.5\textwidth}
  \begin{center}
    \vspace{-0.7cm}
    \includegraphics[width=0.48\textwidth]{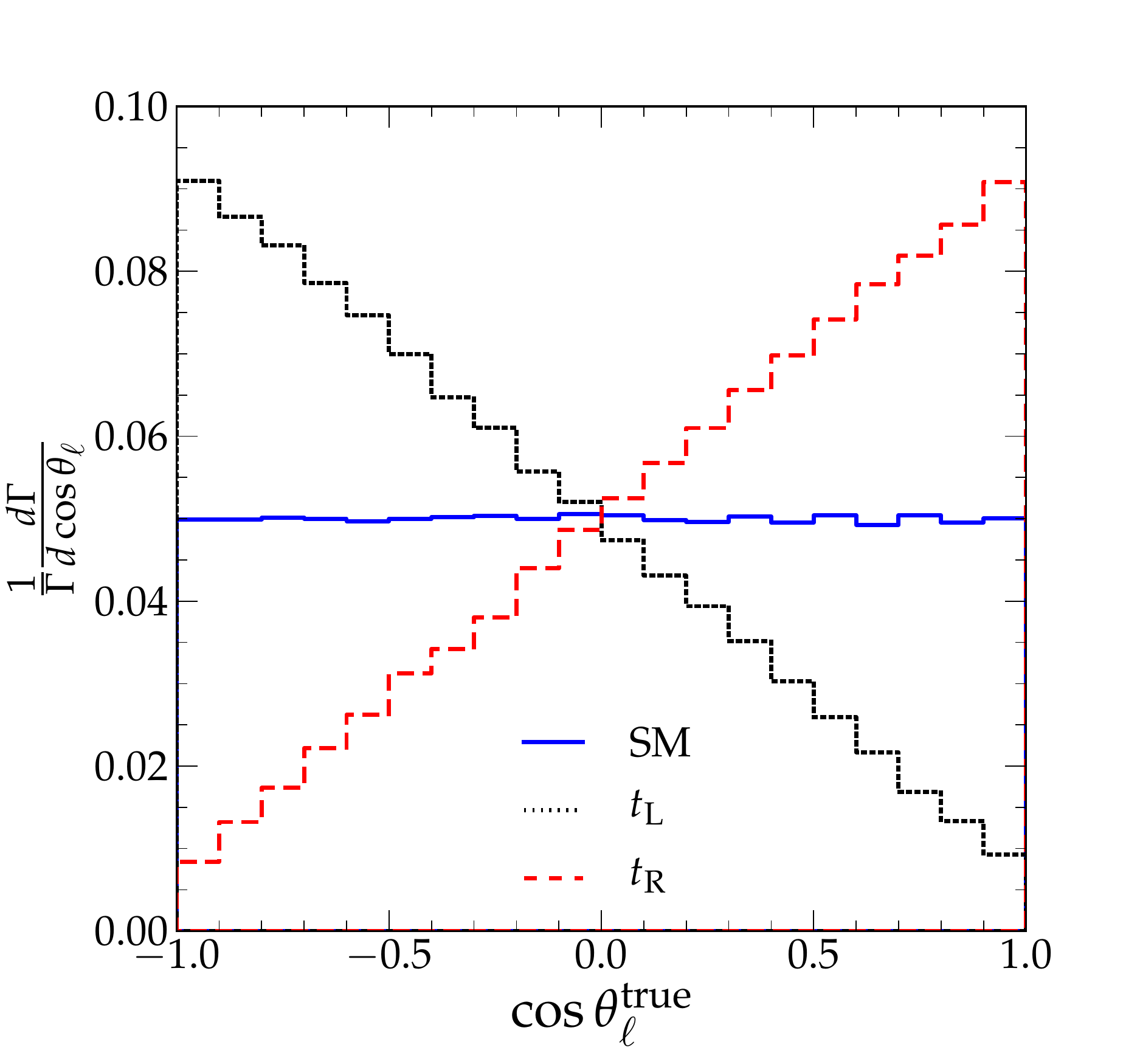}
  \end{center}
  \caption{The distributions of $\cos\theta_\ell$ using the true top-quark momentum.}
  \label{fig:cos_theta_true}
\end{wrapfigure}

By its definition, to calculate the angle $\theta$ one should reconstruct the top rest frame.%
\footnote{%
An alternative strategy is to search for {\em lab-frame} angular observables sensitive to top polarization.
An instance is the lab-frame azimuthal angle of the charged lepton $\phi_\ell$ \cite{Godbole:2006tq}.
For a general analysis of azimuthal-angle distributions, see \cite{BoudjemaSingh}. Yet another approach
is to consider angular variables that depend on longitudinal-boost-invariant combinations of the 
final-state kinematics, such as rapidity differences \cite{Barr2005} (see also 
\cite{MoortgatPick:2011ix}). 
Comparative studies of these variables in the context of new physics can be found in 
\cite{Godbole:2011vw,Edelhauser:2012xb}.}
The $\cos\theta_{\ell}$ distribution in $\ttbar$ production followed by a leptonic decay of both $W$
is shown in Fig.~\ref{fig:cos_theta_true}, where we have used the {\em true} top rest frames. The 
figure shows graphically the very distinct $\cos \theta_\ell$ behavior between the $t_L$ and the $t_R$ 
cases dictated by eq. (\ref{eq:dGamma_dcostheta}).%
\footnote{The figure implicitly includes the anti-top decays as well. This is the case also elsewhere 
in the paper, whenever we do not specify the charge of the lepton.}
\indent Fig.~\ref{fig:cos_theta_true} is purely theoretical, because the two neutrinos in the final 
state challenge the reconstruction of the rest frames of the two tops. Experimentally, event
reconstruction in this case is performed via maximum-likelihood criteria, such as the 
neutrino-weighting method \cite{Abbott:1998dn}, used in \cite{exp_tpol_neutrino}, or 
matrix-element weighting techniques \cite{Chatrchyan:2011nb}, as in \cite{CMS:2012owa} (cf. also
sec. \ref{sec:method_comparison}).

\begin{figure}[tb!]
  \begin{center}
    \includegraphics[width=0.48\textwidth]{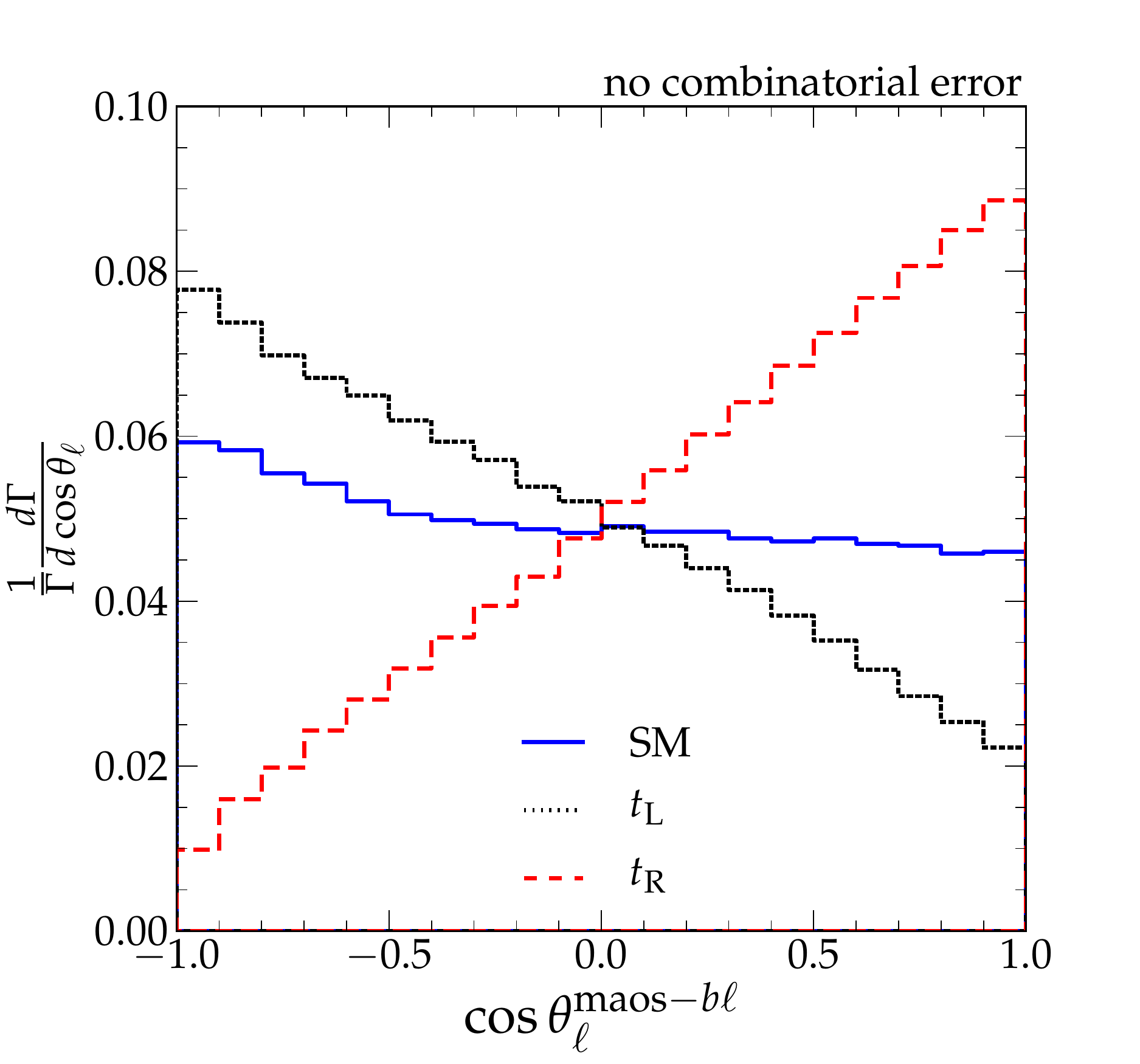}
    \includegraphics[width=0.48\textwidth]{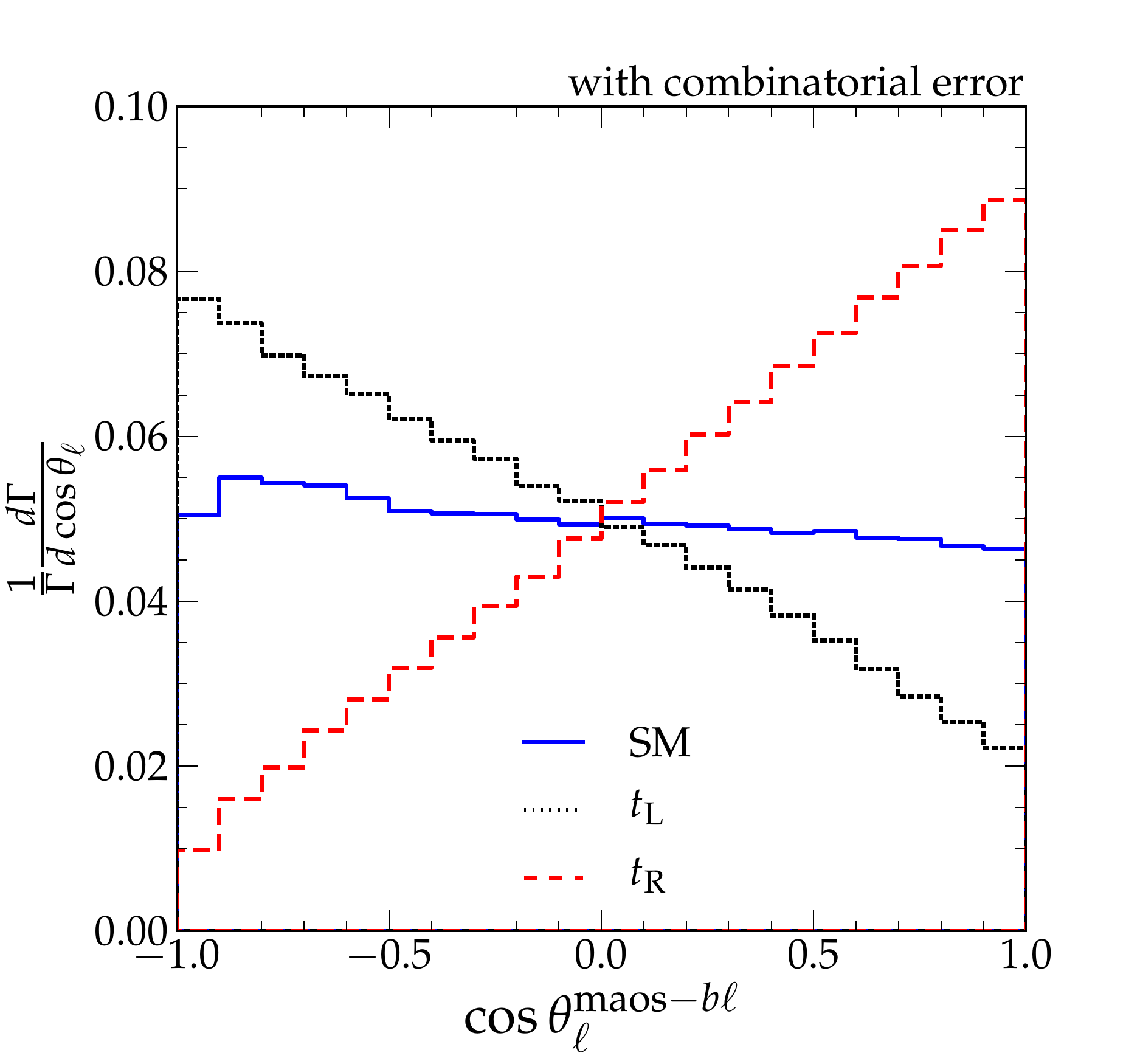}
  \end{center}
  \caption{The distributions of $\cos\theta_\ell$ using the MAOS-reconstructed top-quark momentum 
  without (left panel) and with (right panel) inclusion of the $b\ell$-assignment combinatorial 
  ambiguity (see text).}
  \label{fig:cos_theta_maos_bl}
\end{figure}

We attempt this reconstruction with the MAOS method, and denote the correspondingly calculated angle 
as $\theta_\ell^{\maos}$.%
\footnote{As spin-quantization direction we take the helicity, measured in the $\ttbar$ rest frame.}
Specifically, we again calculate the neutrino momenta from the full-system $\MT2$. Denoting them as
$k^{(i) \maos-b\ell}$, the parent-particle boost is reconstructed event by event as 
$p_{t(\bar t)}^{\maos-b\ell} = p_b + p_\ell + k^{(i)\maos-b\ell}$, with $i = 1,2$ labelling the $t$
or $\bar t$ decay chain. The resulting distributions for purely left-, purely right-handed, and SM 
$\ttbar$ production are shown in the left panel of Fig.~\ref{fig:cos_theta_maos_bl}.

Two observations are in order.
First, the just mentioned reconstruction of the $p_{t (\bar t)}$ momenta suffers
from the combinatorial ambiguity of correctly pairing the two $b$-jets 
%(that are not flavor-tagged, i.e. their charge is not measured in general)
with the two charged leptons.
The left panel of Fig.~\ref{fig:cos_theta_maos_bl} does not include this combinatorial ambiguity --
the $b \ell$ pairings are namely taken to be the correct ones. This plot is meant to show the 
modifications with respect to the true distributions, coming from the MAOS reconstruction alone.
The combinatorial error is included in the right panel of Fig.~\ref{fig:cos_theta_maos_bl}.
This error can be straightforwardly addressed by implementing the four ($\MT2$-based) test 
variables proposed in \cite{Choi_MT2comb}.\footnote{%
For another $\MT2$-based technique to address the same problem see \cite{Baringer_MT2comb}.}
We find that the method correctly assigns the two $b\ell$ pairs in 83\% of the events, before 
any cut. Henceforth, we will refer to the method's percentage of events with correctly assigned 
pairs as efficiency.

A second observation concerns the (only) non-negligible distortion of the MAOS-reconstruct\-ed
distributions with respect to the truth-level ones. This distortion, as apparent from Fig. 
\ref{fig:cos_theta_maos_bl}, occurs for leptons produced in the backward direction 
($\cos \theta_{\ell} \approx -1$), that is well populated in the $t_L$ and SM cases. 
We have investigated in detail the origin of this distortion. A first general explication is the
fact that this region is inherently unfavorable for the application of the MAOS algorithm. In fact,
leptons produced backwards with respect to the parent tops have an energy spectrum peaked towards
softer values (see e.g. Fig.~\ref{fig:x_l_true}, left), whereas the MAOS-algorithm reliability 
increases with larger  visible momenta, as detailed in sec. \ref{sec:MAOS-dilep-tt}. 
Another, more technical, reason for the distortion is the fact 
that kinematic configurations with one of the visible daughter particles produced backwards 
with respect to the parent tend more often to have an `unbalanced' $\MT2$ value 
\cite{Barr:2003rg,Lester:2007fq,Cho:2007dh}. (Namely the $\k{\T}^{(1)}$, $\k{\T}^{(2)}$ 
configuration yielding $\MT2$ is such that $\MT2 = \max\{M_{\T}^{(1)}, M_{\T}^{(2)}\}$, with 
$M_{\T}^{(1)} \neq M_{\T}^{(2)}$. On the other hand, in a balanced solution one has by definition 
$\MT2 = M_{\T}^{(1)} = M_{\T}^{(2)}$.)
Invisible momenta reconstructed from unbalanced $\MT2$ solutions are more likely to deviate
from the true momenta than if they come from balanced $\MT2$ solutions.\footnote{%
\label{foot:balanced_MT2}
This statement is easy to understand for endpoint events, where $\MT2 = m_t$. If $\MT2$ is balanced,
then $m_t = \MT2 = M_{\T}^{(1)} = M_{\T}^{(2)}$, and the uniqueness of the minimum will guarantee
that the MAOS momenta for both decay chains will correspond to the true momenta. On the other
hand, if $\MT2$ is unbalanced, and taking for definiteness $M_{\T}^{(1)} > M_{\T}^{(2)}$, then 
$m_t = \MT2 = M_{\T}^{(1)} \neq M_{\T}^{(2)}$, so that only the MAOS momentum for the first decay 
chain will be the true one, whereas the MAOS momentum for the second decay chain will in general 
deviate from the true momentum.}
As a check, we have repeated Fig.~\ref{fig:cos_theta_maos_bl} (left), but excluding 
events with unbalanced $\MT2$ solutions, and indeed the distortion gets mildened.

\begin{figure}[tb!]
  \begin{center}
    \includegraphics[width=0.48\textwidth]{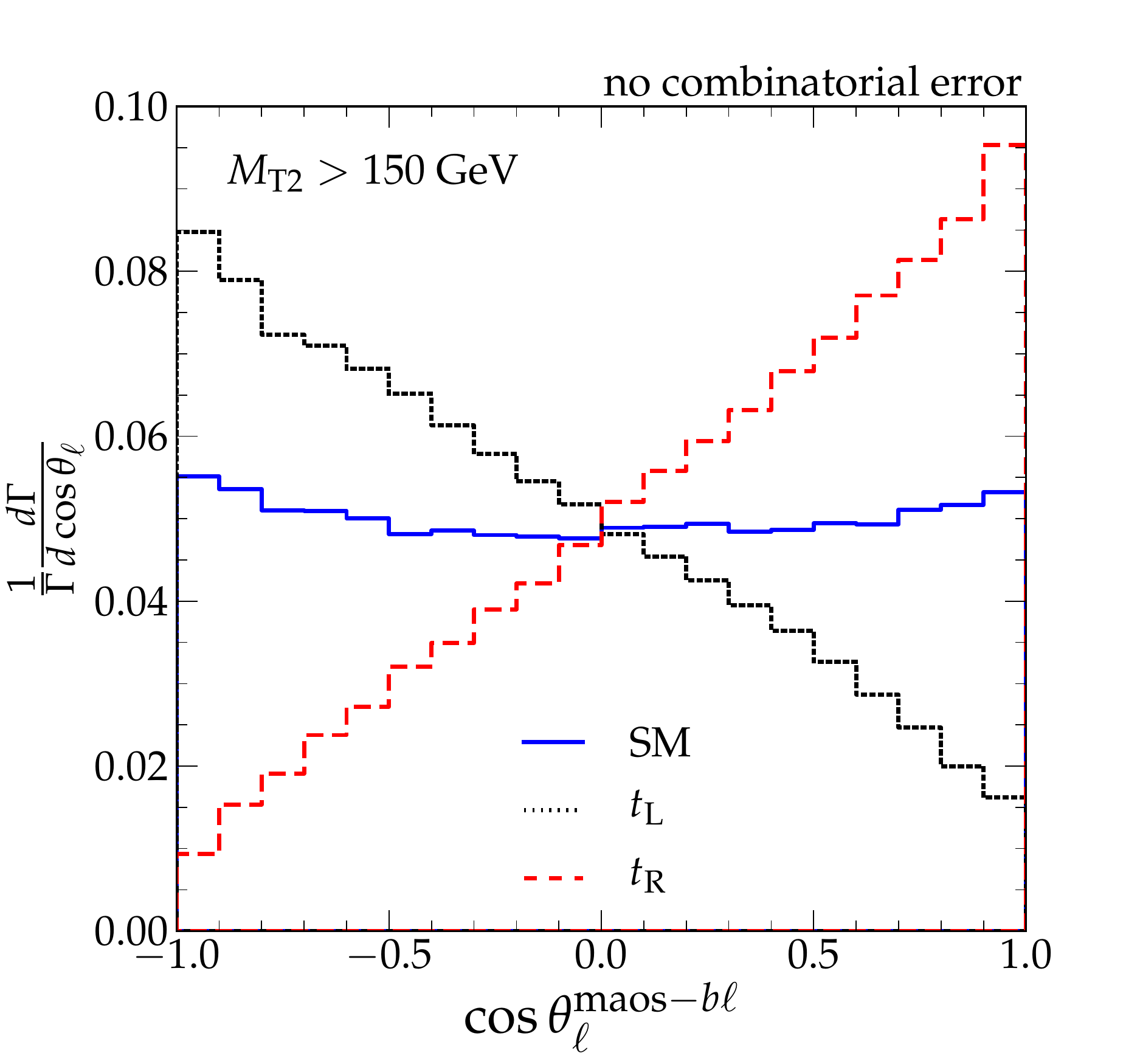}
    \includegraphics[width=0.48\textwidth]{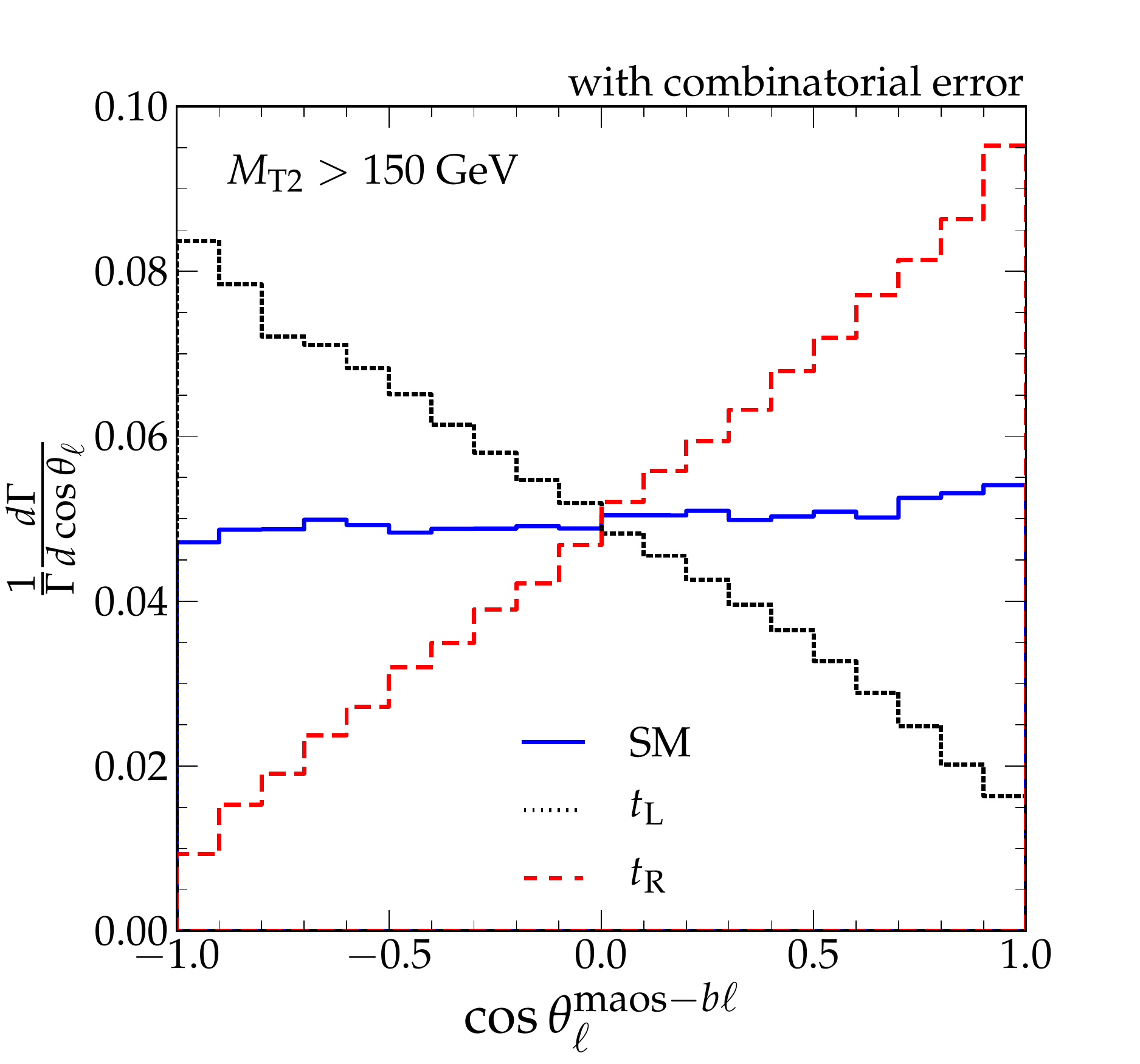}
  \end{center}
  \caption{Same as Fig.~\ref{fig:cos_theta_maos_bl}, but for the inclusion of an $\MT2$ cut.}
  \label{fig:cos_theta_maos_bl_w_cut}
\end{figure}

Both of these effects -- the combinatorial ambiguity and the $\cos \theta_{\ell}^+ \approx -1$
distortion -- can be systematically improved by selecting events with $\MT2$ closer to its endpoint,
where incidentally the MAOS algorithm itself is known \cite{MAOS} to be more reliable 
-- see lower-left panel of Fig.~\ref{fig:delta_kt_mt2}. Furthermore, a lower cut on $\MT2$ represents 
a standard cut in detector-level analyses. In Fig. \ref{fig:cos_theta_maos_bl_w_cut} we show 
histograms which differ from those in Fig.~\ref{fig:cos_theta_maos_bl} for the application of an 
$\MT2 > 150$ GeV cut, that halves the number of events. The figure demonstrates how the cut indeed 
effects positively both the MAOS method alone (left panel) as well as the MAOS method with 
combinatorial error included (right panel). It should also be noted that in the right panel of 
Fig. \ref{fig:cos_theta_maos_bl_w_cut} the above-discussed distortion has largely disappeared. 

All in all, from the sequence of plots in Figs. 
\ref{fig:cos_theta_maos_bl} - \ref{fig:cos_theta_maos_bl_w_cut} one can draw three non-trivial 
conclusions:
{\em (i)} the MAOS reconstruction of the invisible momenta is accurate enough for 
top-polarization distributions not to be appreciably distorted with respect to the true ones; 
{\em (ii)} the combinatorial error (intrinsic to the method, at least for the full-system $\MT2$
case) has in fact a marginal impact on the MAOS reconstruction;
{\em (iii)} this error can be systematically controlled by the same sort of cuts that also help 
the MAOS method itself.

As a side comment to item {\em iii}, we note that in fact the most effective cut to improve the 
efficiency of the combinatorial method is a lower cut on the full-system transverse mass 
$M_{\T}^{\ttbar}$ (see ref. \cite{Choi_MT2comb} for quantitative details). The interesting aspect 
is that this variable is correlated with the overall boost of the $\ttbar$ system -- the harder 
the $M_{\T}^{\ttbar}$ cut, the more boosted the selected $\ttbar$ sample. Therefore, we expect 
our method to perform very well also in the boosted regime.\footnote{An intuitive argument for this 
is the fact that, for boosted tops, a wrong pairing leads very frequently to kinematic solutions 
outside of the physical boundaries, and this information can be exploited to take the other pairing 
as the correct one.}

\begin{figure}[tb!]
  \begin{center}
    \includegraphics[width=0.48\textwidth]{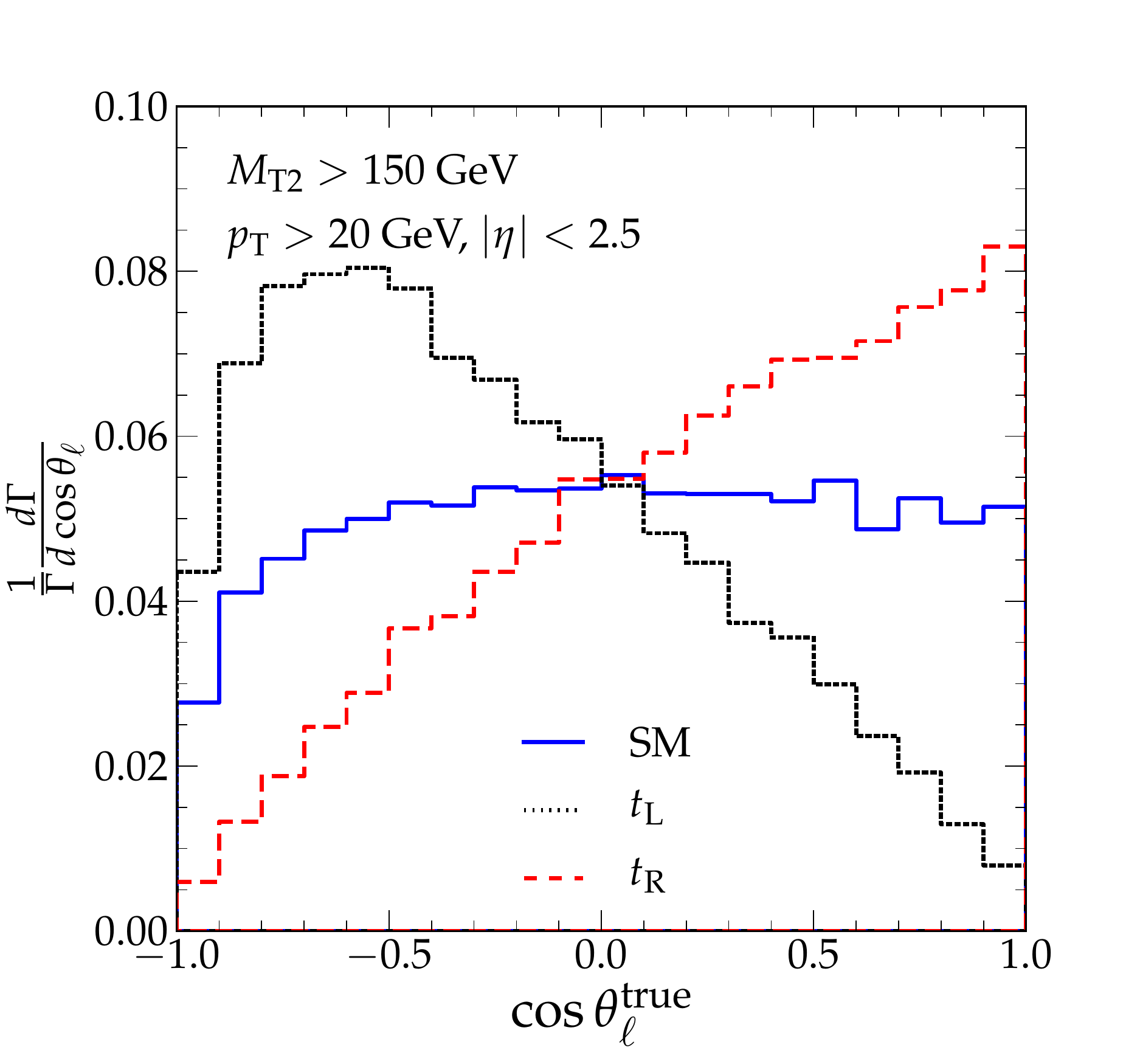}
    \includegraphics[width=0.48\textwidth]{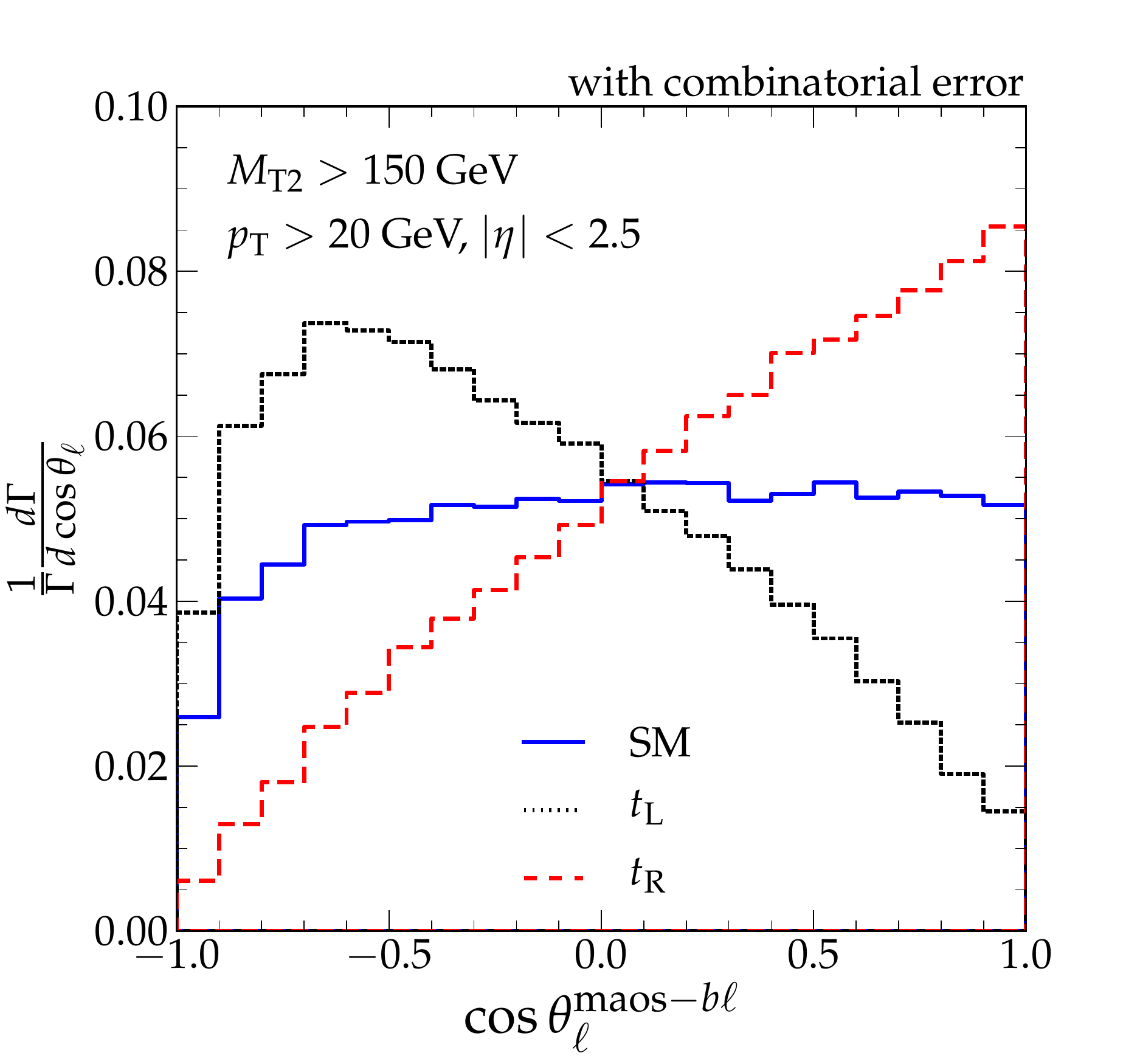}
  \end{center}
  \caption{Comparison between the true (cf. Fig. \ref{fig:cos_theta_true}) and the MAOS-reconstructed 
  (cf. right panel of Fig. \ref{fig:cos_theta_maos_bl_w_cut}) distributions, but for the inclusion 
  of a $p_{\T}$ and an $|\eta|$ cut. In the true distribution (left panel) we also include an $\MT2$ 
  cut for consistency with the analysis of the MAOS-reconstruced distribution (right panel).}
  \label{fig:cos_theta_true_vs_maos}
\end{figure}

A more realistic comparison between the truth-level and the MAOS-reconstructed distributions would
involve the inclusion of a set of cuts such that the selected event sample resemble as much
as possible the one selected by the experimental trigger, as well as the simulation of hadronization 
and energy-momentum smearing effects. We refrain from a refined analysis of this kind in a theory 
study. Similarly as in \cite{MahlonParke1997}, we limit ourselves to the introduction of two `minimal'
centrality cuts, that approximately identify the kinematic fiducial region of the $\ttbar$ sample.
From the recent Atlas analysis \cite{ATLAS:2012ao}, we conservatively take these cuts to be 
$p_{\T} > 20$ GeV and $|\eta| < 2.5$, applied to all final states.\footnote{This choice is rather 
qualitative also because the analysis \cite{ATLAS:2012ao} refers to 7 TeV data. We think nonetheless
that for our main line of argument it is sufficient to use an approximate, conservative figure.} 
We show in Fig. \ref{fig:cos_theta_true_vs_maos} how the true distribution (cf. Fig. 
\ref {fig:cos_theta_true}) and the MAOS-reconstructed one (cf. right panel of Fig. 
\ref {fig:cos_theta_maos_bl_w_cut}) are modified by the introduction of these cuts. As expected, the 
main effect is to underpopulate the bins with $\cos \theta_\ell$ close to $-1$, where the charged 
lepton tends to be softer, as already discussed. Noteworthy is that this distortion effects the true 
and the MAOS-reconstructed distribution in a very similar way.

A more quantitative idea of the difference between all the discussed cases may be obtained by 
calculating the asymmetry observable
\be
\mathcal{A_\ell} = \frac{\sigma (\cos\theta_\ell > 0) - \sigma
(\cos\theta_\ell < 0)}{\sigma (\cos\theta_\ell > 0) + \sigma
(\cos\theta_\ell < 0)}~.
\label{eq:A_ell}    
\ee
Table \ref{tab:A_ell} collects the values of this asymmetry, calculated for the true distribution vs.
the MAOS-reconstructed one for purely left-handed, purely right-handed, or SM-produced $\ttbar$ pairs
(table columns) and without or with inclusion of the most significant cuts and effects discussed
(table rows). To limit clutter, we have labelled the considered cases by the corresponding figure.
These cases include, in order of descending row, the following distributions: 
{\em (i)} the true one; {\em (ii)} the MAOS-reconstructed one, without combinatorial error or cuts; 
{\em (iii)} the MAOS one, including combinatorial error and the $\MT2$ cut;
{\em (iv)} the true one, with $p_{\T}$ and $\eta$ cuts included;  
{\em (v)} the MAOS one as in item {\em iii}, with $p_{\T}$ and $\eta$ cuts included.
As previously discussed in detail, the most relevant comparisons are between cases {\em (i)} and 
{\em (iii)} and between cases {\em (iv)} and {\em (v)}.

\begin{table}[t]
\begin{center}
\begin{tabular}{l|ccc}
                      & $t_L$ & $t_R$ & SM \\ 
\hline
\hline
$\mathcal A_\ell^{\true \phantom{s}}$ (Fig. \ref{fig:cos_theta_true}) 
  & $-0.43$ & $0.43$ & $-0.001$ \\
[0.1cm]
$\mathcal A_\ell^\maos$ (Fig. \ref{fig:cos_theta_maos_bl}, left) 
  & $-0.27$ & $0.41$ & $-0.05$ \\
[0.1cm]
$\mathcal A_\ell^\maos$ (Fig. \ref{fig:cos_theta_maos_bl_w_cut}, right) 
  & $-0.33$ & $0.42$ & $0.03$ \\
[0.1cm]
$\mathcal A_\ell^{\true \phantom{s}}$ (Fig. \ref{fig:cos_theta_true_vs_maos}, left) 
  & $-0.37$ & $0.38$ & $0.05$ \\
[0.1cm]
$\mathcal A_\ell^\maos$ (Fig. \ref{fig:cos_theta_true_vs_maos}, right) 
  & $-0.28$ & $0.40$ & $0.07$ \\
\end{tabular}
\end{center}
\caption{Numerical comparison of the MAOS-reconstructed vs. truth-level $\mathcal A_\ell$ as defined 
in eq. (\ref{eq:A_ell}). The considered cases (table rows) are labelled by the corresponding figure.
The $\ttbar$ pairs are assumed to be purely left-handed, purely right-handed, or produced via SM
QCD (table columns).}
\label{tab:A_ell}
\end{table}

\subsection{Remarks on the method's comparison with existing ones} \label{sec:method_comparison}

Having introduced all the main method's features in a concrete application, it is worthwhile to
comment at this point on how ours compares to existing methods aimed at the reconstruction
of the $t$ and $\bar t$ boosts in di-leptonic $\ttbar$ decays. As already mentioned
in sec. \ref{sec:tpol_ang}, these `likelihood-based' methods include the neutrino weighting ($\nu$W) 
method \cite{Abbott:1998dn} as well as the matrix-element weighting (MEW) one \cite{Chatrchyan:2011nb}.

Each of these methods involves as a crucial step the construction of weighting functions 
(based on Monte Carlo procedures) to estimate the likelihood of each of the possible solutions for 
the neutrino momenta compatible with the system of kinematic equations.

We advance the following remarks on the various methods.

\begin{itemize}

\item Both methods, ours and the likelihood methods, are kinematics-based: in the likelihood methods 
one solves a system of equations corresponding to all kinematic constraints; in ours one uses an $\MT2$ 
property that somewhat summarizes these very kinematic constraints.

\item One difference is in the treatment of the resulting kinematic solutions. We expect that the 
robustness of the $\nu$W and MEW methods will depend on the reliability of the Monte Carlo used to 
determine the weighting function. On the other hand, our method relies solely on kinematics, namely 
on the decay topology being suitable for the construction of $\MT2$.

\item Still concerning the kinematic information used, we further remark that likelihood methods use 
both of the $m_W$ and $m_t$ constraints, that actually differ event by event due to the finite $W$ and 
$t$ widths. On the other hand, the MAOS method is using only one of these mass constraints, $m_t$ in the 
case of $\MT2^{b\ell}$. The explicit use of less kinematic information may be beneficial to reduce the 
associated systematic uncertainty.

\item The accuracy of the MAOS method can be controlled by an $\MT2$ cut, as long as the statistics 
permits it, as is expectedly the case at LHC14. We are not aware of a systematic and intuitive way 
to control accuracy in likelihood methods.

\end{itemize}
In general, it is to be expected that the information from one method will improve the efficiency 
of the other ones (and vice versa). Therefore, barring strong correlations across the methods, the 
best overall reconstruction efficiency will be obtained by integrating (and optimizing accordingly) 
the MAOS method with the other ones.

\section{\boldmath Spin correlations in $\ttbar$ production}\label{sec:spin_corr}

Top polarization may be unobservable if its production mechanism involves $t_L$ and
$t_R$ in similar fractions. This is the case in the SM, where tops are produced
dominantly as $\ttbar$ pairs by QCD interactions, which weigh equally left-handed and right-handed
components. The SM distribution in Fig.~\ref{fig:cos_theta_true}
is in fact unobservably flat (cf. also top-polarization asymmetries in the last column of Table
\ref{tab:A_ell}). In these circumstances, the different spin components in $\ttbar$ 
production can still be tested by looking at spin correlations between $t$ and $\bar t$. The
latter impart correlations between the angular distribution of decay product $i$ 
from the top and the angular distribution of decay product $\bar i$ from the anti-top. 
The doubly-differential (with respect to these two decay products) distribution can be written 
as \cite{MahlonParke_last}
\be
\label{eq:spin_corr}
\frac{1}{\sigma} \frac{d^2 \sigma}{d \cos \theta_i \, d \cos \theta_{\bar i}} ~=~
\frac{1 \, + \, C_{\ttbar} \, \alpha_i \alpha_{\bar i} \, \cos \theta_i \cos \theta_{\bar i}}{4}~,
\ee
where $\theta_{i(\bar i)}$ is the angle between the chosen spin-quantization axis and
the direction of decay product $i (\bar i)$, viewed in the respective mother-particle's rest 
frame, and $\alpha_{i(\bar i)}$ has already been introduced below eq. (\ref{eq:dGamma_dcostheta}).
Furthermore
\be
\label{eq:Cttbar}
C_{\ttbar} ~=~ 
\frac{\sigma_{\tutu}+\sigma_{\tdtd}-\sigma_{\tutd}-\sigma_{\tdtu}}%
{\sigma_{\tutu}+\sigma_{\tdtd}+\sigma_{\tutd}+\sigma_{\tdtu}}~,
\ee
where the symbols on the r.h.s. denote the cross sections for production of $\ttbar$
pairs in either of the four possible spin configurations, with $\uparrow$ ($\downarrow$)
denoting a particle with spin up (down) with respect to the chosen spin-quantization axis.

Spin correlations within the SM, as well as the question how they can best be measured 
at hadron colliders, have been extensively studied ~\cite{MahlonParke1,StelzerWillenbrock,%
Brandenburg,ParkeShadmi,MahlonParke1997,Bernreuther,Uwer,MahlonParke_last}.
Given the $C_{\ttbar}$ dependence in eq. (\ref{eq:spin_corr}), it is clear that spin correlations 
are larger when $C_{\ttbar}$ increases in magnitude, namely when the difference between like- 
and unlike-spin $\ttbar$ pairs is maximal. One crucial insight by Mahlon, Parke and Shadmi
\cite{MahlonParke1,ParkeShadmi,MahlonParke1997} is the realization that this difference can be 
maximized by an appropriate choice of the spin-quantization axis. Once the appropriate basis choice 
is made, the $\ttbar$ cross section turns out to be dominated by one single spin configuration. 
This `optimal' basis choice is different between the Tevatron and the LHC.

At Tevatron, $\ttbar$ pairs are produced dominantly through $q \bar q$ annihilation. For this process,
it has been shown \cite{ParkeShadmi,MahlonParke1997} that one can choose a spin basis in which the 
like-spin $\ttbar$ components ($\tutu$ and $\tdtd$) in the cross section vanish identically,
and this basis is referred to as the `off-diagonal' basis \cite{ParkeShadmi}. A very useful 
parameterization of the corresponding spin eigenvector is provided by Uwer in \cite{Uwer}. In the 
$\ttbar$ rest frame, the angle between the incoming beam (usually identified with $+\hat z$) and the 
top spin axis reads
\be
\label{eq:psi}
\tan \psi ~=~
\frac{\tan \theta \left( 1 - \gamma^{-1} \right)}{1 + \gamma^{-1} \tan^2 \theta}~,
\ee
with $\theta$ the top-quark scattering angle with respect to $\hat z$ and $\gamma = 1/\sqrt{1-\beta^2}$, 
$\beta$ being the top-quark speed. Note that, close to threshold ($\gamma \rightarrow 1$), the spin 
axis becomes aligned to the beam axis, so that in this limit one recovers the `beam-line' basis 
\cite{MahlonParke1}, whereas at very high energies ($\gamma \gg 1$), the spin axis becomes aligned 
to the top-momentum direction, it namely coincides with the `helicity' basis \cite{MahlonParke1}. Concretely, for $\ttbar$ pairs dominantly produced close to threshold,
as was the case at Tevatron, the off-diagonal basis lies close to the beamline basis for all
scattering angles \cite{ParkeShadmi}. For Tevatron data, this suggests to use the off-diagonal 
basis as the optimal choice for spin-correlation studies, and the beamline basis as a sub-optimal 
choice \cite{ParkeShadmi,MahlonParke1997}.

At the LHC, $\ttbar$ pairs are produced dominantly through $gg$ fusion. For $gg \rightarrow \ttbar$, 
there is no basis where the $\ttbar$ pairs are in purely like- or unlike-spin configurations, because 
this basis is {\em different} depending on whether $gg$ are in like- or unlike-helicity configurations, 
and in general both helicity components are present in the $gg \rightarrow \ttbar$ cross section 
\cite{MahlonParke_last,Uwer}.
An `optimized' spin-basis choice can still be made according to whether the $gg$ pair is dominantly 
in a like- or unlike-helicity configuration \cite{MahlonParke_last}, which in turn 
depends on the center-of-mass energy of the $pp$ collisions, or equivalently on $M_{\ttbar}$. 
Specifically, at low $M_{\ttbar}$, $gg$ pairs are dominantly produced with like helicities. In this 
case, the amplitudes squared yielding $\ttbar$ pairs in unlike-spin configurations can be made to 
vanish (for all $\beta$) in the helicity basis \cite{MahlonParke_last}. Conversely, at (very) high 
$M_{\ttbar}$, $gg \rightarrow \ttbar$ occurs dominantly via unlike-helicity gluons. In this case, 
the amplitudes squared to $\ttbar$ in like-spin configurations vanish in the off-diagonal basis 
\cite{MahlonParke_last}, similarly as in the $q \bar q \rightarrow \ttbar$ case seen in the previous
paragraph.
However, as noted in \cite{MahlonParke_last}, the fraction of $\ttbar$ pairs produced in this 
ultrarelativistic limit at the LHC (with 14 TeV collision energy) is very small.\footnote{%
The cross-over point between $gg$ in like- vs. unlike-helicity configurations occurs for $M_{\ttbar} 
\approx 850$ GeV, and at that point the overall $gg \rightarrow \ttbar$ cross section has decreased 
by more than one order of magnitude \cite{MahlonParke1}.}
One concludes that, at the LHC with 14 TeV, $\ttbar$ spin correlations are well described in the 
helicity basis \cite{MahlonParke1,MahlonParke_last}. We will then use this basis for our reference
study of $\ttbar$ spin correlations reconstructed via the MAOS algorithm. We will afterwards address 
the case where an `optimized' basis is used.

\subsection{MAOS-reconstructed spin correlations in the helicity basis} \label{sec:spin_corr_hel}

From eq. (\ref{eq:spin_corr}) it is clear that, along with an accurate choice of the 
spin-quantization axis, it is also essential to choose correctly the final states $i$ and $\bar i$ 
-- different states have different spin-analyzing power, and pose different detection challenges 
\cite{Jezabek,StelzerWillenbrock}.
In this paper, since we are focusing on kinematic methods based on $\MT2$, we confine ourselves
to $\ttbar \rightarrow b \ell^+ \bar \nu \, \bar b \ell^- \nu$ and take 
$i = \ell^+$ and $\bar i = \ell^-$. As already remarked, di-leptonic $\ttbar$ decays have the advantage
that the charged-leptons' spin-analyzing power is maximal, $|\alpha_{\ell^+}| = |\alpha_{\ell^-}| = 1$, 
and that $\ell^\pm$ are clean objects experimentally, and the drawback of two final-state neutrinos,
that hinder the reconstruction of the $t$ and $\bar t$ rest frames,\footnote{%
An alternative approach to testing $\ttbar$ spin correlations is to look for variables that can
be measured in the lab frame. An example is the difference between the azimuthal angles of the 
two charged leptons in di-leptonic $\ttbar$ decays \cite{MahlonParke_last}. We will not pursue this
approach in the present work.} required by the $\theta_{\ell^{\pm}}$ definition. Experimentally, 
the two main techniques used \cite{Abbott:1998dn,Chatrchyan:2011nb} have already been mentioned in 
the context of top polarization. They have been applied in $\ttbar$ spin correlation studies in 
\cite{exp_spin_corr}.
%\cite{CMS:2012cxa,Abbott:2000dt,Abazov:2011qu,Abazov:2011ka}

In sec. \ref{sec:tpol_ang} on top polarization we have estimated these angular variables
using the MAOS-reconstructed invisible momenta $k^{(i)\maos-b\ell}$. We namely calculated the $t$ or 
$\bar t$ rest frames as $p_{t(\bar t)}^{\maos-b\ell} = p_b + p_\ell + k^{(i)\maos-b\ell}$, using 
the values for the invisible momenta that yield $\MT2$ for the event, and reconstructed the angles 
$\theta_{\ell^\pm}$ accordingly. Here we apply this technique to reconstruct the $\ttbar$ 
spin-correlation in eq. (\ref{eq:spin_corr}).

\begin{wrapfigure}{r}{0.5\textwidth}
  \vspace{-0.3cm}
  \includegraphics[width=0.48\textwidth]{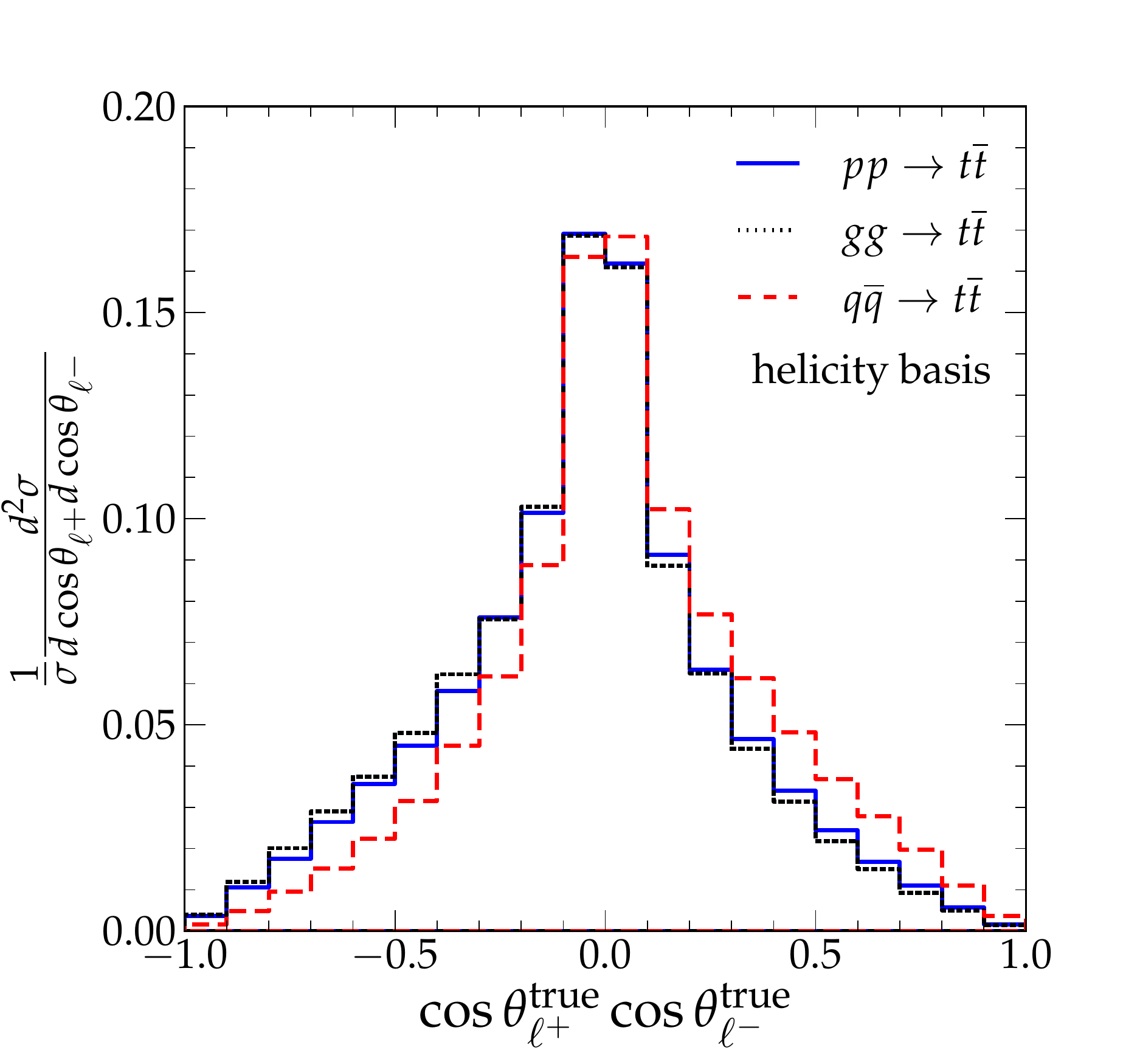}
  \caption{The differential distribution in eq. (\ref{eq:spin_corr}) as a function of 
  $\cos\theta_{\ell^+} \cos\theta_{\ell^-}$ for $\ttbar$ produced from $pp$ collisions
  at 14 TeV and by its subprocesses $gg$ and $q \bar q$ (see legend).
  Spin axes are in the helicity basis.}
  \vspace{-0.0cm}
  \label{fig:costh_corr_true}
\end{wrapfigure}

As in the top-polarization study, we provide truth-level distributions as a reference. A convenient 
quantity to measure the `size' of $\ttbar$ spin correlations from the distribution in eq. 
(\ref{eq:spin_corr}) is the asymmetry 
\cite{StelzerWillenbrock}
\be
A_{\ell \ell} ~ \equiv ~ 
\frac{N(\cos\theta_{\ell^+} \cos\theta_{\ell^-} > 0) - N(\cos\theta_{\ell^+} \cos\theta_{\ell^-} < 0)}%
{N(\cos\theta_{\ell^+} \cos\theta_{\ell^-} > 0) + N(\cos\theta_{\ell^+} \cos\theta_{\ell^-} < 0)}~=
\frac{1}{4} C_{\ttbar} \, \alpha_{\ell^+} \alpha_{\ell^-}~,
\label{eq:All}
\ee
where $N$ denotes the number of events satisfying the condition in parentheses, and we have specialized
the notation to di-leptonic $\ttbar$. This asymmetry may be visualized from the dependence of the 
differential distribution in eq. (\ref{eq:spin_corr}) on the product $\cos \theta_{\ell^+} \, 
\cos \theta_{\ell^-}$. In Fig. \ref{fig:costh_corr_true} we show such dependence in the case of the 
{\em true} distribution of $\ttbar$ produced in $pp$ collisions at 14 TeV (as well as via the 
subprocesses $gg$ and $q \bar q$).\footnote{%
We use the CTEQ6L1 parton distribution functions \cite{PDF}.}
The superscript `true' in the $x$-axis 
emphasizes that in this plot the $t$ and $\bar t$ rest frames are calculated using the true 
neutrino momenta, and also assigning the correct $b \ell$ pairs to the two decay chains, i.e. without 
including combinatorial ambiguities. 
%As spin-quantization axis we choose, as mentioned, the $t$ and $\bar t$ helicities, measured in the 
%$\ttbar$ rest frame \cite{MahlonParke1}.

\begin{figure}[hb]
  \begin{center}
    \includegraphics[width=0.48\textwidth]{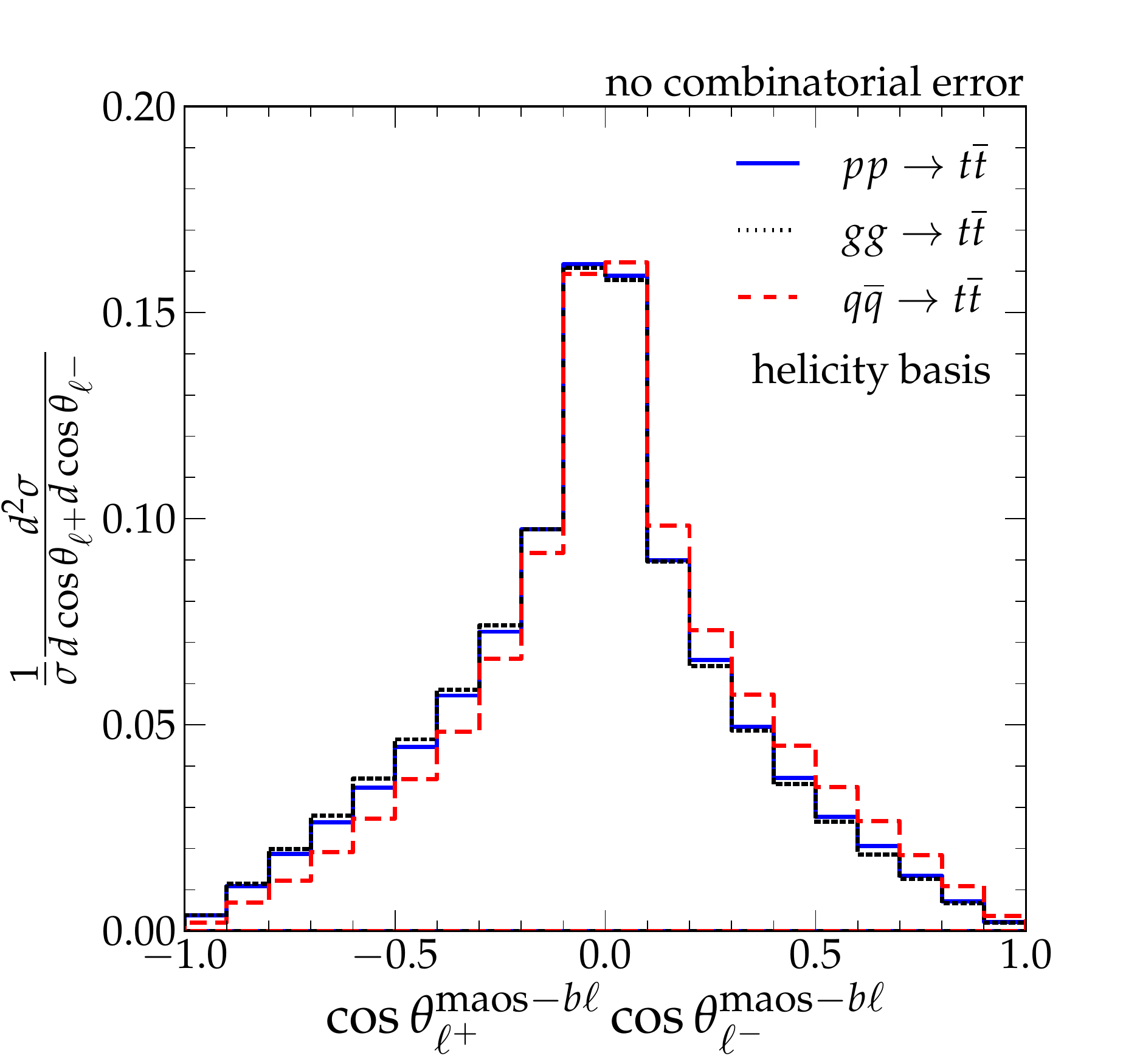}
    \includegraphics[width=0.48\textwidth]{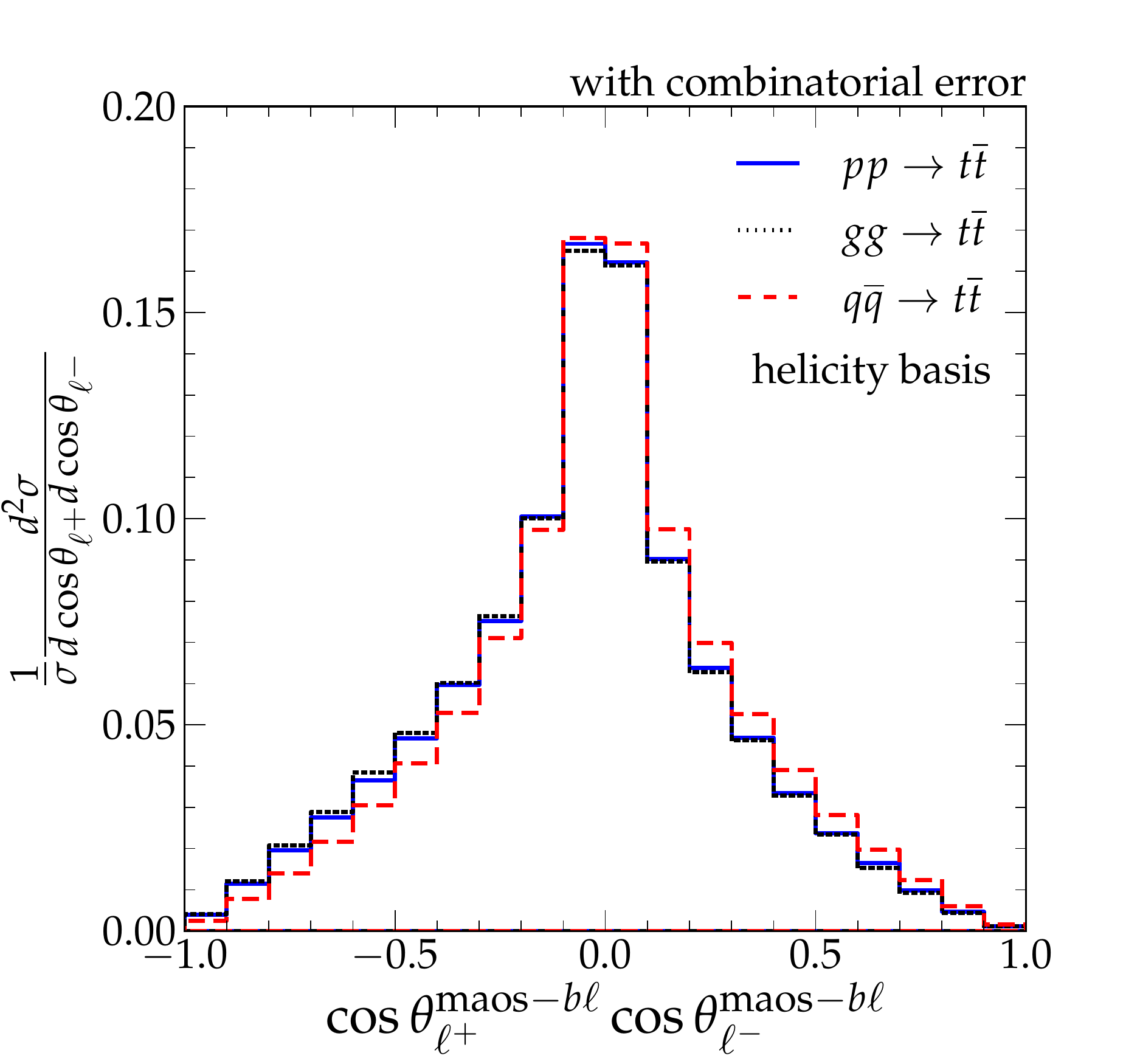}
  \end{center}
  \caption{Same distributions as in Fig. \ref{fig:costh_corr_true}, but with MAOS-reconstructed
  $t$ and $\bar t$ boosts (see text). The left panel does not include the $b\ell$ combinatorial 
  ambiguity, which is instead taken into account in the right panel.}
  \label{fig:costh_corr_maos_bl}
\end{figure}

We now turn to the MAOS-reconstruct\-ed version of the histograms in Fig.~\ref{fig:costh_corr_true}.
The {\em equivalent} histograms, namely including no other uncertainty than the one due to MAOS-estimated 
$t$ and $\bar t$ boosts rather than the true boosts, are shown in the left panel of 
Fig.~\ref{fig:costh_corr_maos_bl}. In the right panel, we also include the combinatorial ambiguity
of assigning the two $b\ell$ pairs to the $t$ and $\bar t$ decay chains. This ambiguity is addressed
via the same method \cite{Choi_MT2comb} as the one used in the top-polarization study, sec. 
\ref{sec:tpol_ang}. Fig. \ref{fig:costh_corr_maos_bl} provides an already non-trivial test: the 
clear-cut asymmetry visible in Fig. \ref{fig:costh_corr_true} is still present in the histogram with 
MAOS-reconstructed momenta and combinatorial ambiguity included.

Inspection of Fig. \ref{fig:costh_corr_maos_bl} reveals that the asymmetry is more pronounced after 
inclusion of the combinatorial error (right panel) than before it (left panel). This is due to the 
fact that the efficiency of the combinatorial method is higher in the $\cos \theta_{\ell^+} 
\cos \theta_{\ell^-} < 0$ region than in the $\cos \theta_{\ell^+} \cos \theta_{\ell^-} > 0$ one: it 
equals 87\% vs. 79\% before cuts. As a consequence, some of the wrongly-paired solutions belonging to 
$\cos \theta_{\ell^+} \cos \theta_{\ell^-} > 0$ will migrate to the other region, whereas the converse 
will happen less likely. So, while the MAOS method ignoring combinatorial ambiguities slightly dilutes 
the asymmetry in Fig. \ref{fig:costh_corr_true}, the inclusion of combinatorial ambiguities largely 
compensates this dilution, yielding an asymmetry closer to the truth-level one.

\begin{figure}[tb!]
  \begin{center}
    \includegraphics[width=0.48\textwidth]{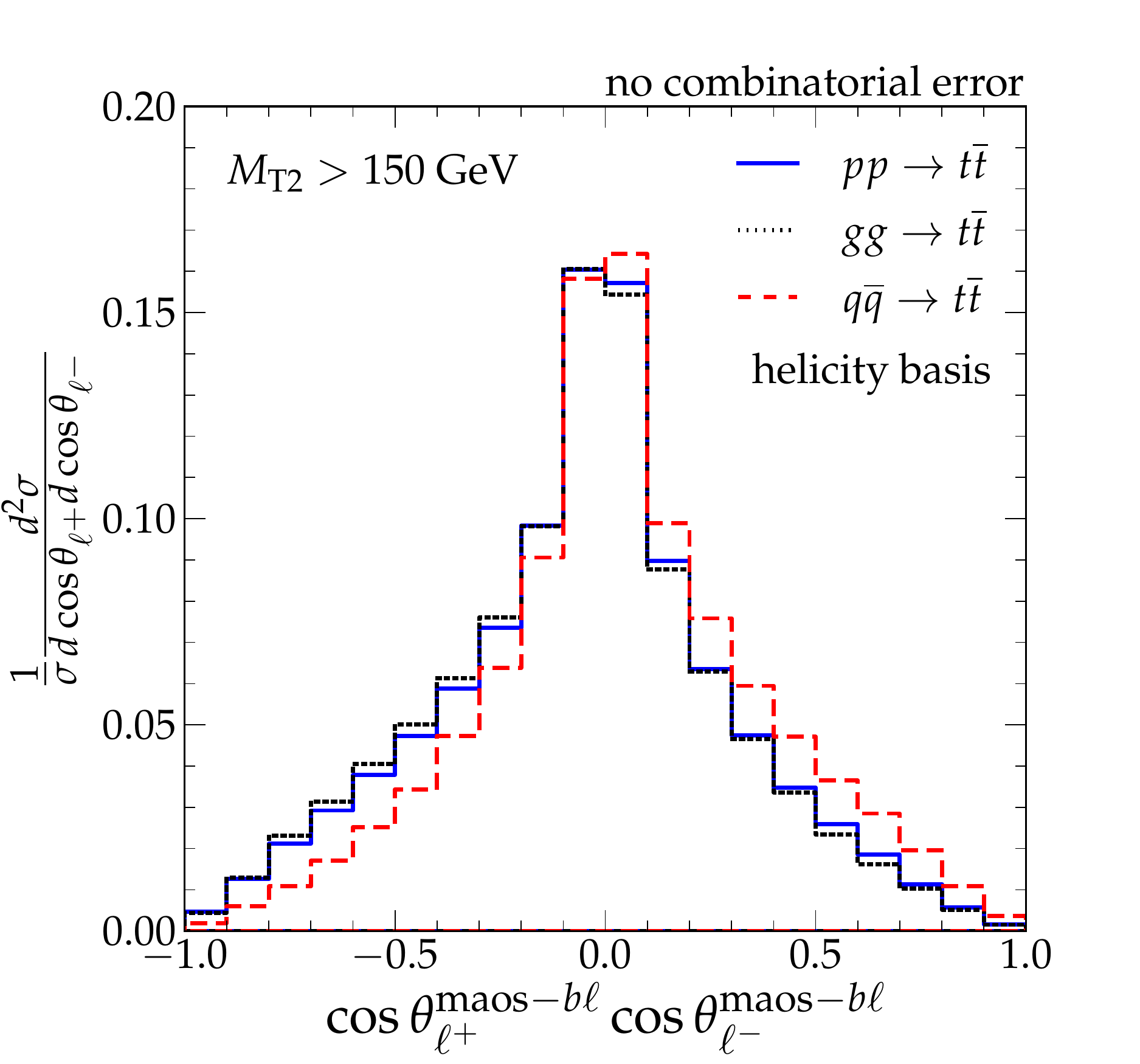}
    \includegraphics[width=0.48\textwidth]{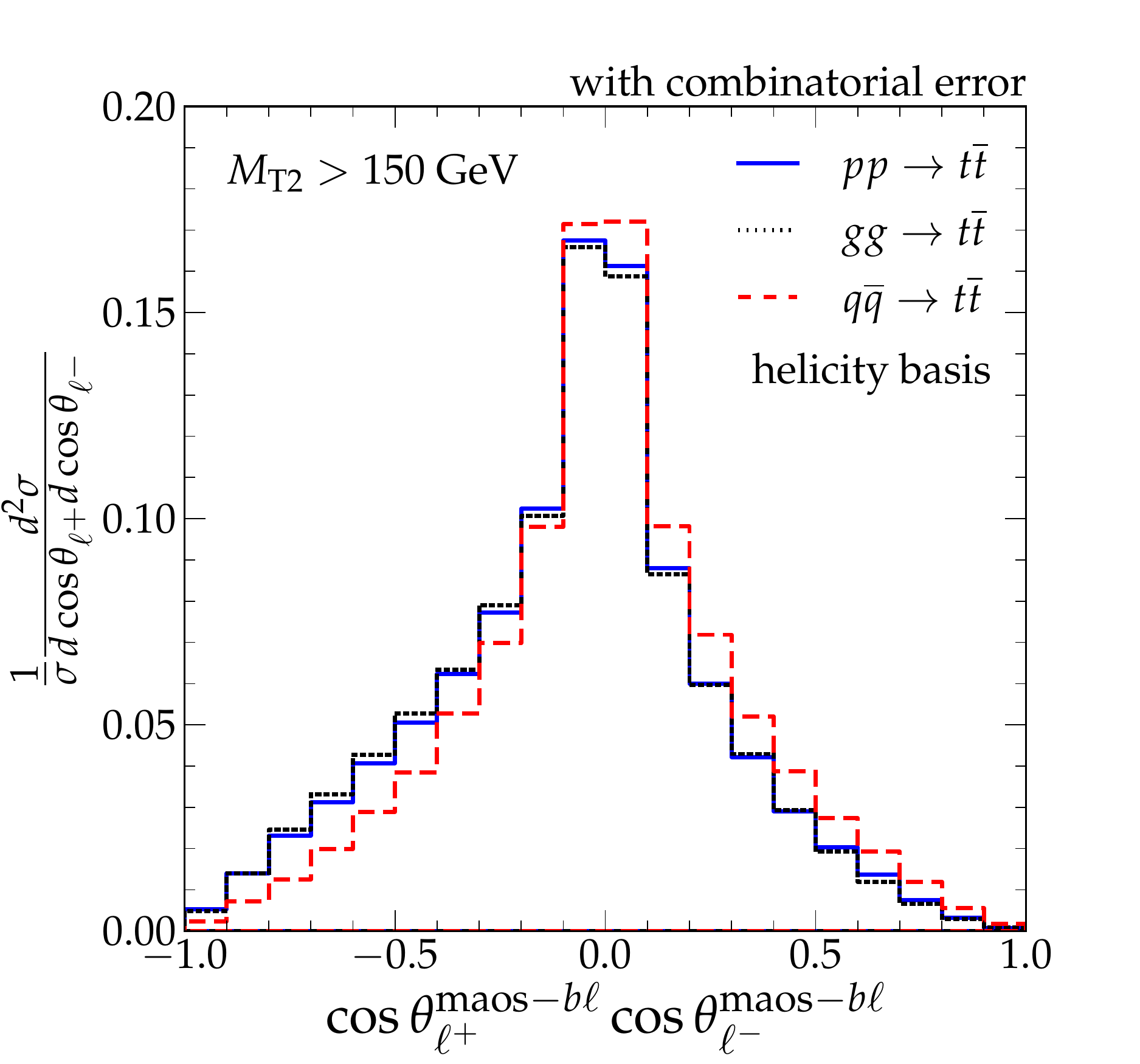}
  \end{center}
  \caption{Same distribution as Fig. \ref{fig:costh_corr_maos_bl}, but for the introduction of 
  an $\MT2 > 150$ GeV cut.}
  \label{fig:costh_corr_maos_bl_cut}
\end{figure}

The spurious asymmetry component induced by the treatment of combinatorial ambiguities can be made 
to disappear by a suitable $\MT2$ cut. In fact, the latter reduces substantially the difference 
between the negative and the positive $x$-axis efficiencies -- with the cut $\MT2 > 150$ GeV, the 
two efficiencies equal 92\% vs. 91\%.
In Fig. \ref{fig:costh_corr_maos_bl_cut} we show the same MAOS distributions as Fig. 
\ref{fig:costh_corr_maos_bl}, but for the inclusion of the requirement $\MT2 > 150$ GeV.
As already discussed, the introduction of an $\MT2$ cut is beneficial to the MAOS-reconstruction 
reliability, and in fact the left panel of Fig. \ref{fig:costh_corr_maos_bl_cut} displays a larger
asymmetry than the corresponding panel of Fig. \ref{fig:costh_corr_maos_bl}. 
Turning to the right panel of Fig. \ref{fig:costh_corr_maos_bl_cut}, where the combinatorial uncertainty
is taken into account, we note that the asymmetry is increased with respect to the case of no cut.
We observe in this respect that, {\em per se}, an $\MT2$ cut has the effect of increasing the 
asymmetry -- already at the level of the true distribution. For example, the asymmetry $A_{\ell \ell}$ 
(\ref{eq:All}) in the $pp$ case equals $-0.087$ for the truth-level distribution of Fig. 
\ref{fig:costh_corr_true} (cf. also Table \ref{tab:All} to follow) and reaches $-0.114$ for the 
same histogram in presence of an $\MT2 > 150$ GeV cut.

\begin{figure}[tb!]
  \begin{center}
    \includegraphics[width=0.48\textwidth]{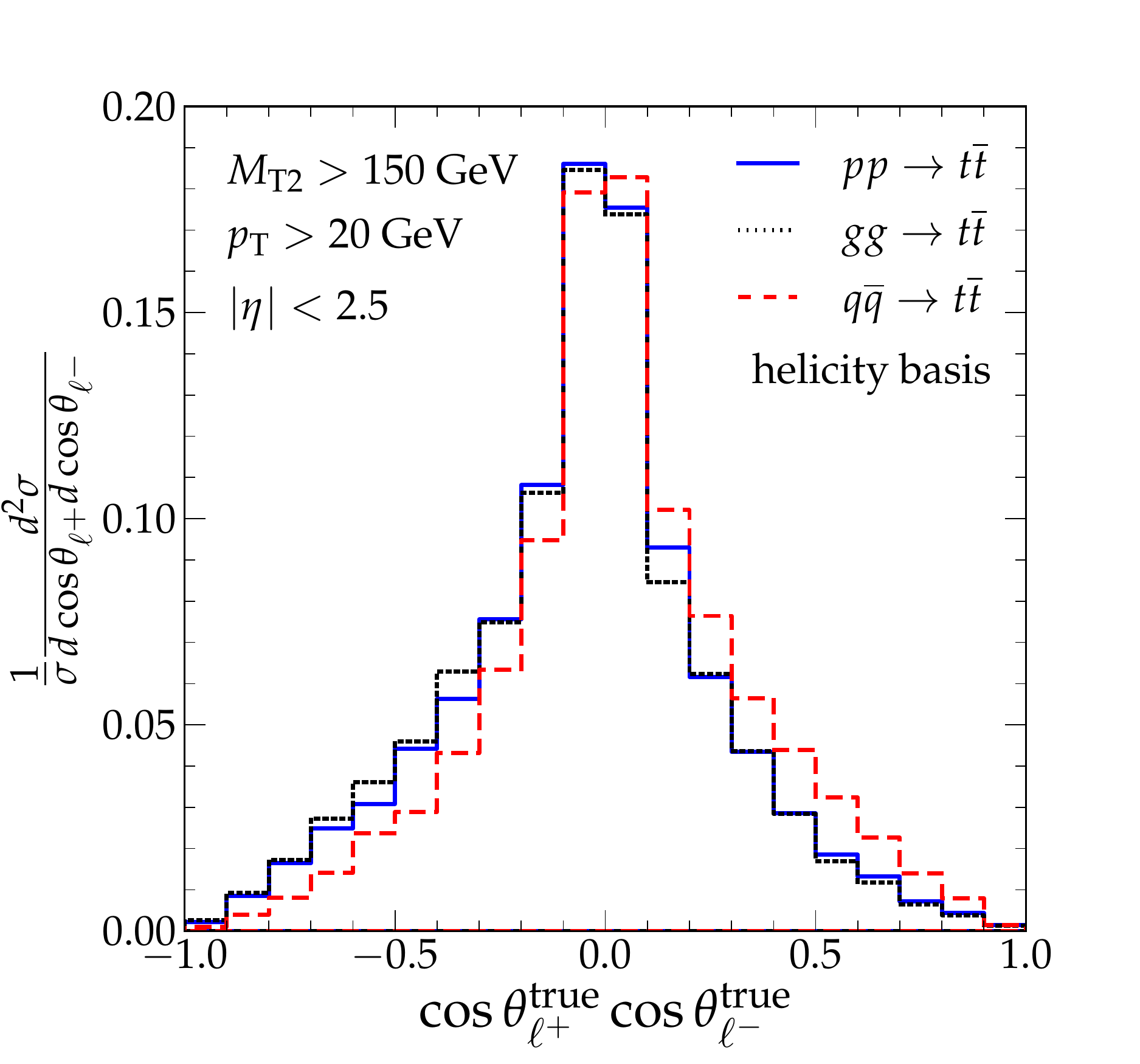}
    \includegraphics[width=0.48\textwidth]{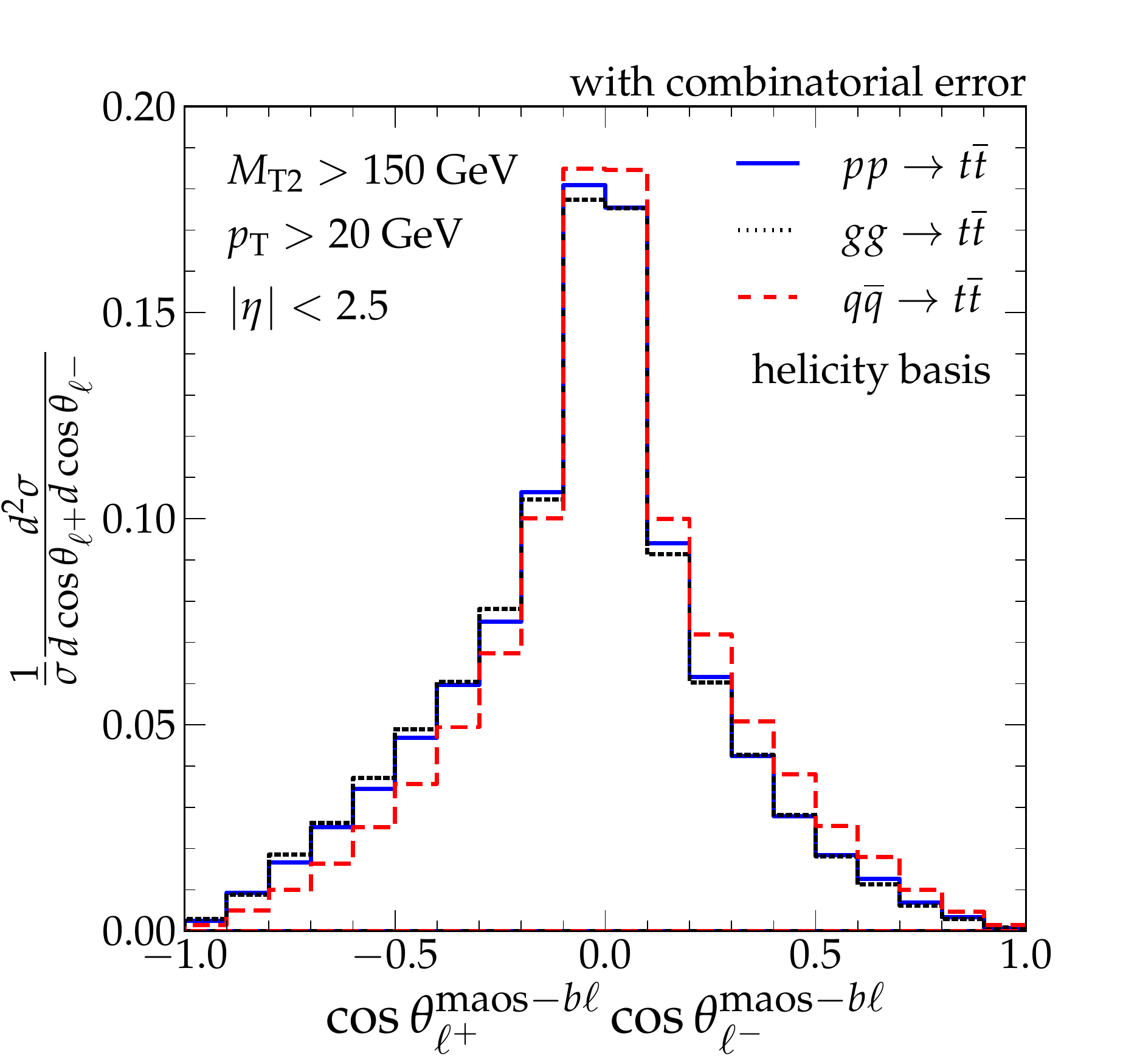}
  \end{center}
  \caption{Comparison between the true (cf. Fig. \ref{fig:costh_corr_true}) and the MAOS-reconstructed 
  (cf. right panel of Fig. \ref{fig:costh_corr_maos_bl_cut}) spin-correlation distributions, but for 
  the inclusion of a $p_{\T}$ and an $|\eta|$ cut. In the true distribution (left panel) we also include 
  an $\MT2$ cut for consistency with the analysis of the MAOS-reconstruced distribution (right panel).}
  \label{fig:costh_corr_true_vs_maos}
\end{figure}

As a final comparison of the MAOS-reconstructed distribution with respect to the truth-level one, 
we include in both cases the two minimal centrality cuts $p_{\T} >$ 20 GeV and $|\eta| < 2.5$, as
already discussed in the top-polarization study (cf. end of sec. \ref{sec:tpol_ang}). The resulting 
distributions are shown in Fig. \ref{fig:costh_corr_true_vs_maos}. Worth remarking is the fact that
the $p_{\T}$ and $|\eta|$ cuts do not introduce major distortions in these distributions. As seen 
in the top-polarization discussion, this sort of cuts is expected to underpopulate the kinematic
region where one of the charged leptons is produced backwards with respect to the parent, 
$\cos \theta_\ell \approx -1$. Since the angle of the other lepton is generic, the effect is diluted
in the whole $\cos \theta_{\ell^+} \cos \theta_{\ell^-} \in [-1, +1]$ range, and does not visibly
affect Fig. \ref{fig:costh_corr_true_vs_maos}.

We conclude this section by calculating the asymmetry parameter $A_{\ell \ell}$ defined in eq. 
(\ref{eq:All}) in the most representative among the cases discussed. These values are collected in
Table \ref{tab:All}. The considered cases include (in order of descending row): {\em (i)} the true
asymmetry, without inclusion of any errors or cuts; {\em (ii)} the corresponding MAOS-reconstructed
asymmetry, again without errors or cuts; {\em (iii)} the MAOS asymmetry, with inclusion of the 
combinatorial ambiguity and of an $\MT2$ cut; {\em (iv)} the true asymmetry, with inclusion of
centrality cuts on $p_{\T}$ and $|\eta|$; {\em (v)} the MAOS asymmetry as in item {\em iii}, and 
including the centrality cuts. The most significant comparisons are between cases {\em (i)} 
and {\em (iii)}, and between cases {\em (iv)} and {\em (v)}.

\begin{table}[t]
\begin{center}
\begin{tabular}{l|ccc}
helicity basis & $pp$ & $gg$ & $q \bar q$ \\ 
\hline
\hline
$A_{\ell \ell}^{\true \phantom{s}}$ (Fig. \ref{fig:costh_corr_true}) 
  & $-0.087$ & $-0.12$ & $0.11$ \\
[0.1cm]
$A_{\ell \ell}^\maos$ (Fig. \ref{fig:costh_corr_maos_bl}, left) 
  & $-0.055$ & $-0.075$ & $0.060$ \\
[0.1cm]
$A_{\ell \ell}^\maos$ (Fig. \ref{fig:costh_corr_maos_bl_cut}, right) 
  & $-0.15$ & $-0.16$ & $-0.003$ \\
[0.1cm]
$A_{\ell \ell}^{\true \phantom{s}}$ (Fig. \ref{fig:costh_corr_true_vs_maos}, left) 
  & $-0.11$ & $-0.13$ & $0.080$ \\
[0.1cm]
$A_{\ell \ell}^\maos$ (Fig. \ref{fig:costh_corr_true_vs_maos}, right) 
  & $-0.11$ & $-0.13$ & $0.009$ \\
\end{tabular}
\end{center}
\caption{Numerical comparison of the MAOS-reconstructed vs. the truth-level spin-correlation asymmetry
$A_{\ell \ell}$ as defined in eq. (\ref{eq:All}) and calculated in the helicity basis. The considered 
cases (table rows) are labelled by the corresponding figure. The initial states (table columns) are 
$pp$ at 14 TeV, or its $gg$ or $q \bar q$ subprocesses.}
\label{tab:All}
\end{table}

\subsection{MAOS-reconstructed spin correlations in a boost-dependent basis} \label{sec:spin_corr_opt}

As discussed at the beginning of sec. \ref{sec:spin_corr}, unlike the case of 
$q \bar q \rightarrow \ttbar$ one cannot define an optimal basis to calculate $\ttbar$ spin
correlations in the case of $gg \rightarrow \ttbar$ \cite{MahlonParke_last,Uwer}, because the optimal 
basis is different for like- or unlike-helicity $gg$, and both helicity components are present for $gg$ 
colliding via $pp$ pairs. In practice though, the relative weights of the different helicity components
change with the collision energy, and one may define a $\sqrt s$-dependent spin-quantization basis 
according to the helicity component that is dominant at that $\sqrt s$. A numerical approach to this
possibility was presented in \cite{Uwer}, and an analytic solution in \cite{MahlonParke_last}. 
This paper identifies the relation $\beta \gamma \sin \theta = 1$\footnote{%
With $\beta$ the $t$ boost and $\theta$ its production angle with respect to the beam axis, in the 
rest frame of the colliding partons.}
as the kinematic condition separating the region where like-helicity $gg$ dominate ($\beta \gamma 
\sin \theta \ll 1$) from the one where unlike-helicity $gg$ do ($\beta \gamma \sin \theta \gg 1$). 
Then, in the first (second) region one can define a $\beta$- and $\theta$-dependent spin-quantization 
axis that maximizes the $\tutu + \tdtd$ ($\tutd + \tdtu$) fractions. We henceforth refer to this axis 
as $\psi_{\rm like}$ ($\psi_{\rm unlike}$), in the notation of eq. (\ref{eq:psi}).
As a practical approximation to this basis, Ref. \cite{MahlonParke_last} suggests to use the 
helicity (off-diagonal) basis in the $\beta \gamma \sin \theta < 1$ ($> 1$) region. This suggestion 
can be understood by noting that the region $\beta \gamma \sin \theta \ll 1$ ($\gg 1$) can be 
identified with the near-threshold (ultra-relativistic) regime, and by recalling that, at the LHC, 
the helicity basis performs well near threshold, while the off-diagonal basis does so in the 
ultra-relativistic regime (cf. beginning of sec. \ref{sec:spin_corr}).

\begin{table}[t]
\begin{center}
\begin{tabular}{l|ccc}
hybrid basis & $pp$ & $gg$ & $q \bar q$ \\ 
\hline
\hline
$A_{\ell \ell}^{\true \phantom{s}}$ (Fig. \ref{fig:costh_corr_true}) 
  & $-0.072$ & $-0.11$ & $0.13$ \\
[0.1cm]
$A_{\ell \ell}^\maos$ (Fig. \ref{fig:costh_corr_maos_bl}, left) 
  & $-0.047$ & $-0.067$ & $0.073$ \\
[0.1cm]
$A_{\ell \ell}^\maos$ (Fig. \ref{fig:costh_corr_maos_bl_cut}, right) 
  & $-0.13$ & $-0.15$ & $0.030$ \\
[0.1cm]
$A_{\ell \ell}^{\true \phantom{s}}$ (Fig. \ref{fig:costh_corr_true_vs_maos}, left) 
  & $-0.078$ & $-0.11$ & $0.13$ \\
[0.1cm]
$A_{\ell \ell}^\maos$ (Fig. \ref{fig:costh_corr_true_vs_maos}, right) 
  & $-0.095$ & $-0.11$ & $0.052$ \\
\end{tabular}
\end{center}
\caption{Same as Table \ref{tab:All}, but for the use of the hybrid basis (see text) in
place of the helicity basis.}
\label{tab:All_hyb}
\end{table}
We have repeated the helicity-basis study of sec. \ref{sec:spin_corr_hel} in the basis suggestion
of Ref. \cite{MahlonParke_last}.
We will henceforth refer to this choice as the `hybrid' basis. Our results for $A_{\ell \ell}$ in 
this basis are collected in Table \ref{tab:All_hyb}. By comparing this table with Table \ref{tab:All} 
one immediately notes that the hybrid basis indeed improves the $q \bar q$ component of our 
spin-correlation asymmetries, but it slightly worsens the $gg$ component, which is however the dominant 
one in $pp$ collisions. As a consequence, we find the helicity basis \cite{MahlonParke1} to performs 
globally better than the hybrid basis.

A few comments on these findings are in order.
First, we note explicitly that, by construction, the hybrid basis coincides with the helicity basis
for $\beta \gamma \sin \theta < 1$, including the near-threshold region. By looking at the 
$pp \rightarrow \ttbar$ production cross section at 14 TeV, one easily realizes that the overwhelming 
majority of $\ttbar$ pairs is produced in this region. From this argument alone, it is clear that any 
difference between the helicity and the hybrid basis will affect only the tail of the 14 TeV $\ttbar$ 
distribution.
A second observation concerns the $\beta \gamma \sin \theta > 1$ region, where the hybrid basis becomes 
the off-diagonal one. In fact, it is worth remarking that, for $\beta \gamma \sin \theta > 1$ the 
off-diagonal basis tends analytically to the optimal basis, indicated above by $\psi_{\rm unlike}$, only
for $\cos \theta \rightarrow 0$ \cite{MahlonParke_last}. Away from this limit, deviations between the
two bases occur, and it is not obvious how these deviations affect a given observable.
In this respect, it should be noted that, while for $\beta \gamma \sin \theta < 1$ like-helicity
$gg$ pairs clearly dominate the cross section ($gg_{\rm like} : gg_{\rm unlike} = 55\% : 20\%$),
for $\beta \gamma \sin \theta > 1$ unlike-helicity $gg$ pairs dominate the cross section only slightly
($gg_{\rm unlike} : gg_{\rm like} = 15\% : 10\%$) \cite{MahlonParke_last}.

Our finding that the helicity basis performs somewhat better than the hybrid one for the 
spin-correlation asymmetry (\ref{eq:All}) is specific to SM $pp \rightarrow \ttbar$ production 
at 14 TeV. At higher collision energies and in presence of new physics, the hybrid basis may be
substantially more advantageous for $\ttbar$ spin-correlation studies. We leave this topic outside 
the scope of the present work.

\section{Conclusions}

A known challenge in pair-production of two particles each decaying semi-invisibly is the reconstruction
of the full event kinematics. This reconstruction would on the other hand be very useful: for instance, 
it would be instrumental to testing differential distributions with respect to suitable final-state 
momenta. These distributions would in turn allow to determine the spin fractions with which the decaying
particles are produced, thereby dissecting their production mechanism.

In this paper we have explored this general idea in the benchmark case of $\ttbar$ production followed 
by a leptonic decay of both $W$ bosons -- in this case the two invisible particles are the two 
neutrinos. We have studied the possibility of reconstructing the full $t$ and $\bar t$ boosts using 
the invisible momenta that correspond to the $\MT2$ minimum -- in the literature known as MAOS 
invisible momenta. 

The relevant question is whether the thus reconstructed $t$ and $\bar t$ momenta are faithful enough 
to the true momenta. `Enough' depends in general on the class of observables considered. We test the 
MAOS-reconstructed $t$ and $\bar t$ momenta against observables sensitive to top polarization and 
$\ttbar$ spin correlations, most notably angular distributions of the daughter charged leptons. We find
that the MAOS-reconstructed distributions and the corresponding asymmetries are always very close to 
the truth-level ones, and that the method's performance can be systematically improved by only an 
$\MT2$ cut.

The discussion in this work is confined to $\ttbar$ production from $pp$ collisions at 14 TeV. 
Nonetheless, the main line of argument is clearly applicable to any decay process where one can define 
and calculate $\MT2$, e.g. pair production of new states, each one decaying to visibles plus an 
escaping\footnote{Or treated as such, see comment in footnote \ref{foot:undetected}, page 
\pageref{foot:undetected}.} particle.

In this application, the method would open the possibility of measuring the spin fractions of the 
produced new states, arguably one of the most direct ways to probe the details of the production 
mechanism. Not committing here to any specific model beyond the SM, we leave this direction to future 
work.

\subsection*{Acknowledgments}

DG would like to thank Claude Duhr, and in a special way Benjamin Fuks, for feedback on the use 
of the program \textsc{FeynRules}. The authors would also like to thank Kiwoon Choi, Adam
Falkowski and Yeong Gyun Kim for discussions on topics related to the subject of this work. CBP 
gratefully acknowledges the hospitality of LAPTh Annecy, where part of this work was carried out. 
The work of CBP is supported by the CERN-Korea fellowship through the National Research Foundation 
of Korea.

\bibliographystyle{JHEP}
\bibliography{maos_spin}

\end{document}